\documentclass[%
 twocolumn,
 amsmath,amssymb,
 aps,
prstab,
floatfix,
longbibliography
]{revtex4-1}

\usepackage{graphicx}
\usepackage{dcolumn}
\usepackage{bm}
\usepackage{xcolor}
\usepackage{hhline}
\usepackage{siunitx}
\usepackage[normalem]{ulem}
\usepackage[]{hyperref}
\hypersetup{
    colorlinks=true,
    linkcolor=blue,
    citecolor=blue,
    filecolor=blue,      
    urlcolor=blue,
}

\begin{document}

\preprint{APS/123-QED}

\title{Classification of vacuum arc breakdowns in a Pulsed DC System}

\author{Anton Saressalo}
\email{anton.saressalo@helsinki.fi}
\author{Andreas Kyritsakis}
\author{Flyura Djurabekova}%
\affiliation{%
Helsinki Institute of Physics and Department of Physics, University of Helsinki,
PO Box 43 (Pietari Kalmin katu 2), 00014 Helsingin yliopisto, Finland
}

\author{Iaroslava Profatilova}%
\author{Jan Paszkiewicz}%
\author{Sergio Calatroni}
\author{Walter Wuensch}%
\affiliation{%
CERN, European Organization for Nuclear Research, 1211 Geneva, Switzerland
}




\date{\today}

\begin{abstract}
\noindent
Understanding the microscopic phenomena behind vacuum arc ignition and generation is crucial for being able to control the breakdown rate, thus improving the effectiveness of many high-voltage applications where frequent breakdowns limit the operation. In this work, statistical properties of various aspects of breakdown, such as the number of pulses between breakdowns, breakdown locations and crater sizes are studied independently with almost identical Pulsed DC Systems at the University of Helsinki and in CERN. In high-gradient experiments, copper electrodes with parallel plate capacitor geometry, undergo thousands of breakdowns. The results support the classification of the events into primary and secondary breakdowns, based on the distance and number of pulses between two breakdowns. Primary events follow a power law on the log--log scale with the slope $\alpha \approx \num{1.30}$, while the secondaries are highly dependent on the pulsing parameters.
\end{abstract}

\pacs{Valid PACS appear here}
\maketitle


\section{\label{sec:intro}Introduction}
Grasping the underlying physical processes leading to electrical vacuum arc outbursts -- breakdowns (BDs) -- is important for many applications across various fields in modern science and technology. The phenomenon occurs in devices that operate in (ultra) high vacuum and which are subject to high electric fields. Applications include vacuum switches and interrupters, vacuum arc metal processing, ion beam and pulsed sources, fusion reactors, satellites and radio-frequency (RF) particle accelerators \cite{Boxman1996HandbookApplications,Latham1995HighPractice,Falkingham2004AConnection,McCracken1980ATokamaks}.

Investigation on the origin of BDs has been underway for more than a century \cite{Boxman1996HandbookApplications}. Numerous experimental, theoretical and, more recently, computational studies have been performed over the decades in order to understand the phenomenon \cite{Dyke1953TheInitiation,Charbonnier1967ElectricalTheory,Schade2003NumericalField}. Several different processes have been suggested to explain the arc formation, but none of them have provided adequate analytic explanation \cite{Zhou2019DirectResolution}.

It is clear that surface electric fields below \si{\giga\volt/\meter} range are not strong enough to break a metal surface in order to initiate the plasma of the vacuum arc. The most common explanation is that micro and nanoscale protrusions on the surface locally increase the surface electric field and this enhanced field results in electron field emission induced evaporation of neutral atoms as well. The emitted electrons accelerated under the electric field ionize some of the atoms which in turn are accelerated back towards the cathode, sputtering more neutrals into the vacuum and starting an avalanche process \cite{Barengolts2018MechanismStructures,Kyritsakis2018ThermalEmission}. Leading to an exponential growth in the number of charged particles in the vacuum and practically short circuiting the anode and cathode, this whole process is seen as a vacuum discharge -- vacuum arc -- also known as a breakdown. Once the system is short circuited, large currents flow through even with relatively small voltages. Origin and experimental observation of these protrusions is unclear. Some hypotheses link them to near-surface dislocations causing deformations on the surface \cite{Nordlund2012DefectFields,Engelberg2018StochasticFields}.

The Compact Linear Collider (CLIC) is an example of an application in which breakdowns play an important role in limiting the design. The project is a high-energy physics facility  proposed to be built at CERN in order to accelerate and collide electrons and positrons \cite{Burrows2018TheReport}. The accelerating structures for the main beam of CLIC operate at room temperature and use \SI{50}{\mega\watt} X-band RF pulses to accelerate electrons and positrons in an ultra-high vacuum environment. In order to minimize the length and construction costs of the facility, electric fields up to \SI{100}{\mega\volt/\meter} are used to accelerate the particles for the highest collision energy stage. These high accelerating fields correspond to surface electric fields in excess of \SI[per-mode = symbol]{200}{\mega\volt\per\meter}, a value limited by vacuum electrical breakdowns. If a breakdown occurs during the operation of the CLIC accelerator, the particle beam is kicked and no $e^+e^-$ collisions can occur for that pulse. Thus, the accelerator's luminosity is reduced, which is why the breakdown rate (BDR) is required to be kept below  \num{3e-7} per pulse per meter for the accelerator to operate efficiently \cite{Aicheler2012AReport}.

Copper has been chosen as the  material of these accelerating structures \cite{Zennaro2008DesignStructures}. Despite its relatively low average breakdown field after conditioning, \SI{170}{\mega\volt/\meter}, the other properties of copper, such as good conductivity, machinability, ductility and availability made it the best choice for the material \cite{Descoeudres2009DCVacuum}. 

In order to optimize the accelerating structure design and operation, the structures are experimentally tested in klystron-based X-band test facilities in CERN \cite{Catalan-Lasheras2014ExperienceCERN,CatalanLasheras2016CommissioningStand,Volpi2018HighKlystrons}. These test stands allow \SI{200}{\nano\second} pulses with output up to \SI{50}{\mega\watt} and repetition rate up to \SI{400}{\hertz} with breakdown behaviour being one of the most important parameters investigated \cite{Shipman2014ExperimentalCLIC,Woolley2015HighCavity}. 

Since building and operating these RF test facilities is both expensive and time-consuming, DC pulse experiments have been designed specifically to study the breakdowns with much higher repetition rates and simpler setup. In spite of the differences between the RF and DC systems, the DC pulsing is made as close as possible to the RF case. This way the DC experiments are not only useful in studying the BD resistance during CLIC-like pulsing, but also in understanding the basic physics behind the BD initiation  general.

\section{\label{sec:exp}Experimental setup}
\subsection{Equipment}
The experiments were conducted using a Pulsed DC System aimed to emulate the RF pulsing, but with a higher repetition rate. The system includes two parallel copper electrodes inside a vacuum chamber (Large Electrode System, LES), connected to a high-voltage power supply and a pulse generator along with an oscilloscope and a measurement computer. Almost identical Pulsed DC Systems in CERN and at the University of Helsinki were used in this work. In CERN, there are also continuous experiments done with the actual CLIC accelerating structures, using \SI{12}{\giga\hertz} RF pulses with repetition rates up to \SI{400}{\hertz}.

\begin{figure}
\centering
\includegraphics[width=\linewidth]{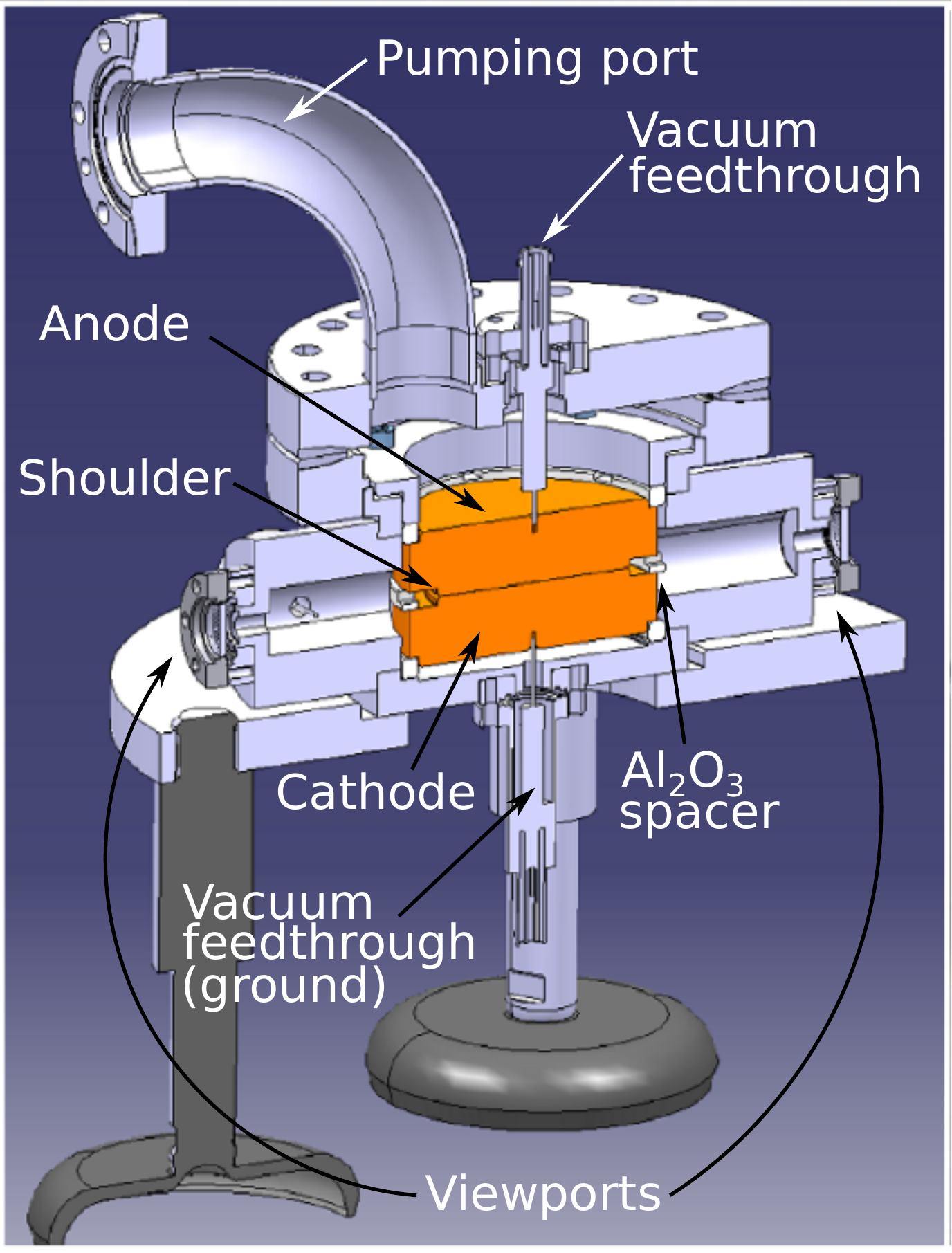}
\caption{\label{fig:LES1Chamber} A cross-section from a 3D model of the LES vacuum chamber. The illustration shows the electrodes in the middle, connected to vacuum feedthroughs and otherwise insulated by an Al$_2$O$_3$ spacer. Four viewports on the sides of the chamber allow also visual detection of a breakdown.}
\end{figure}

LES is a compact vacuum chamber designed at CERN for high-gradient studies (Figure \ref{fig:LES1Chamber}). Inside the chamber, there are two diamond-machined cylindrical electrodes separated by an aluminum oxide spacer which maintains a desired gap between the electrode surfaces. In these studies, spacers resulting in a \SI{60}{\micro\meter} gap were used, but there are also other options for a gap from \SIrange{20}{100}{\micro\meter}. The surface roughness and dimensional precision of the spacers as well as the electrodes is below \SI{1}{\micro\meter} by design.  Together, these electrodes act as a parallel plate capacitor. One of the electrodes is charged to a positive voltage (anode) while the other one is grounded (cathode). The small gap allows generation of electric fields of even \SI{100}{\mega\volt/\meter} with DC voltages of only a few kilovolts. During measurements, the chamber is pumped down to high-vacuum below 10$^{-7}$\,\si{\milli\bar}  using a turbo pump in series with a roughing pump.

High-voltage microsecond-pulses with a repetition rate of typically \numrange[range-phrase = --]{1}{2}\,\si{\kilo\hertz} are generated using a Marx Generator EPULSUS\textsuperscript{\textregistered}-FPM1-10 by Energy Pulse Systems \cite{2015ManualEPULSUS-FPM1-10}. The generator utilizes SiC MOSFET technology for amplifying the voltage from a power supply by a factor of at least \num{10}, depending on the model, and enabling pulses with lengths from \SI{200}{\nano\second} upwards. During a pulse, the effective capacitor inside the LES is charged with a current spike, and a similar spike in the opposite direction discharges the electrodes after a specified up-time (pulse length), provided that no breakdown occurred. The rise and fall time of the pulses are in the order of \SI{100}{\nano\second} \cite{Redondo2017Solid-stateStudies}. Pulse length and repetition rate can be varied programmatically. 

\begin{figure}[!htbp]
\centering
\includegraphics[width=\linewidth]{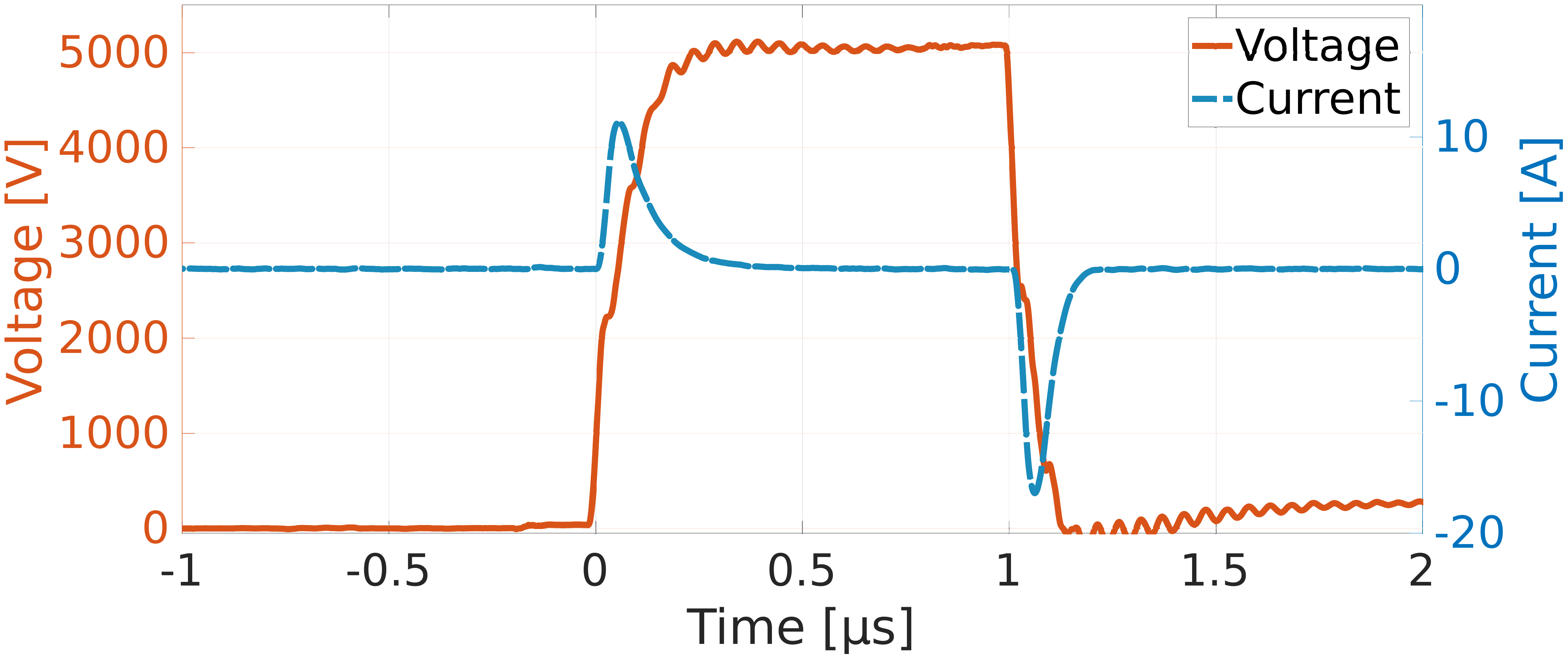}
\caption{\label{fig:WF_noBD} A typical waveform of a \SI{1}{\micro\second} pulse generated by Marx generator without a breakdown. The figure shows the charging and discharging current peaks at the start and at the end the pulse, with  the voltage staying constant in between.}
\end{figure}

\begin{figure}[!htbp]
\centering
\includegraphics[width=\linewidth]{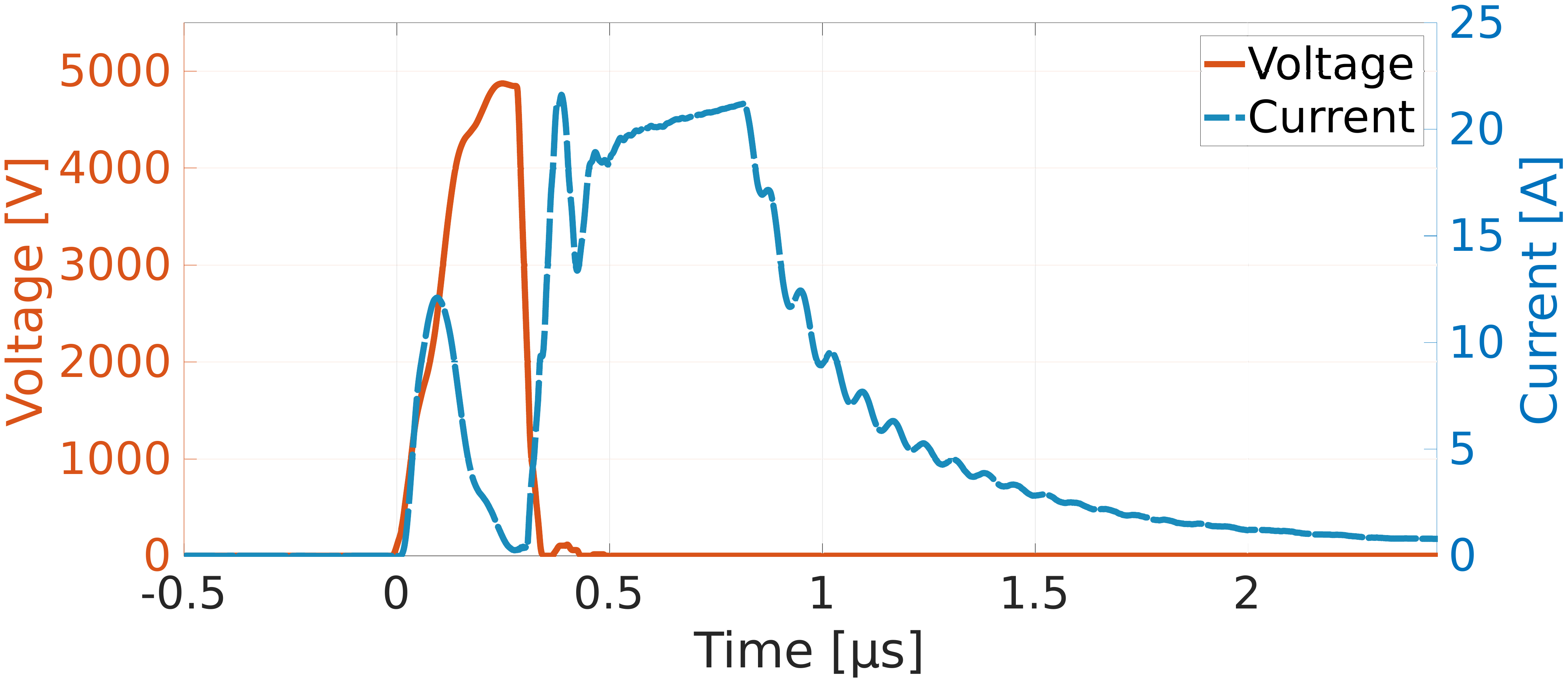}
\caption{\label{fig:WF_BD} A typical waveform of a pulse with a breakdown, showing the extra current peak followed by roughly \SI{600}{\nano\second} of burning voltage. During the breakdown, the voltage drop occurs within approximately five nanoseconds.}
\end{figure}

The generator is also used for detecting the electrical breakdowns between the electrodes by monitoring the current during pulsing. When a breakdown occurs, there is a rapid current peak as the anode and cathode are briefly short circuited. This current peak is typically at least by a factor of \num{2} higher than the charging peak and can thus be distinguished by the generator. After the breakdown peak, the short circuit stays open and there will be a constant \SI{20}{\volt} burning voltage of across the gap for around \numrange[range-phrase = --]{250}{400}\,\si{\nano\second} \cite{Juttner2001CathodeArcs, Shipman2012MeasurementTime}. Examples of waveforms without and with a breakdown in Figures \ref{fig:WF_noBD} and \ref{fig:WF_BD}. Sampling rate for these waveforms in the oscilloscope is \SI{1}{\giga\hertz}.

\subsection{Pulsing \& conditioning algorithm}
The ultimate aim of the experiments is to increase the breakdown resistance of the electrodes during pulsing with as high electric field as possible while keeping the breakdown rate within predefined limits. The increase in the breakdown resistance is achieved by conditioning the copper electrodes with electric pulses \cite{Descoeudres2009DCVacuum}. For copper, this conditioning typically requires a great number of pulses (typically more than $10^{7}$ \cite{Volpi2018HighKlystrons}), during which numerous breakdowns occur (typically more than \num{100}) \cite{Cox1974VariationVacuo}. 

The conditioning starts with relatively low electric fields, for example, \SI{10}{\mega\volt/\meter} and the voltage is gradually increased in small steps after each pulsing period, of typically \num{100000} pulses. If a breakdown occurs, the pulsing for that period is terminated and the electric field is either slightly decreased or not changed at all, depending on the number of pulses which took place until the breakdown. The algorithm is similar to that used in the RF experiments and is explained in details in \cite{Korsback2019VacuumSystem}.

\begin{figure}
\centering
\includegraphics[width=\linewidth]{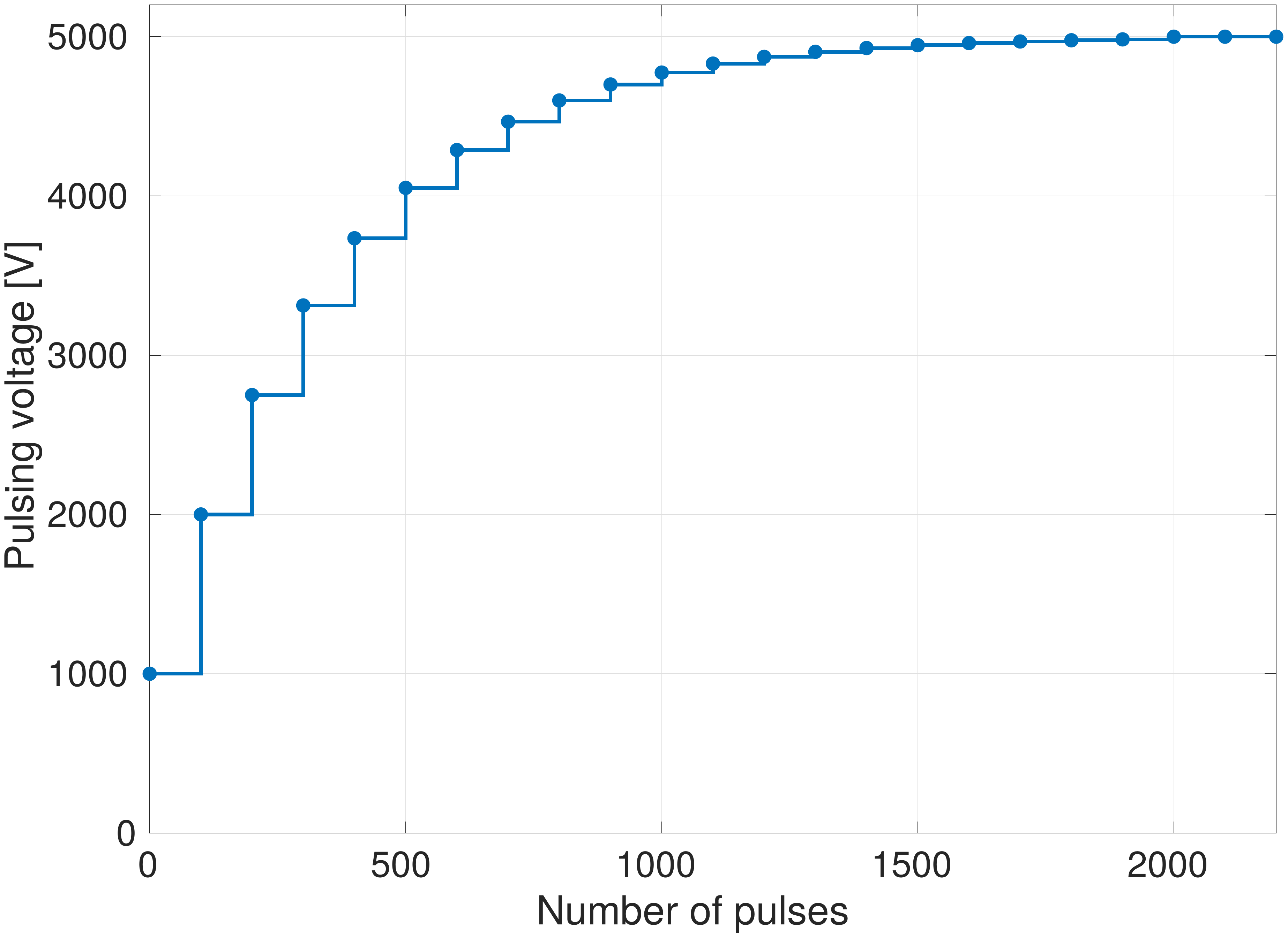}
\caption{\label{fig:ramping} Asymptotic ramping period after a BD at $V=\SI{5000}{\volt}$ (which is $E=\SI{83}{\mega\volt/\meter}$, based on $E=V/d$). Each dot represents the start of a new ramping step of \num{100} pulses. The ramping starts from one fifth of the target voltage. A BD may also occur at any time during the ramping, after which the ramping starts over again, though the target voltage stays the same as before.}
\end{figure}

During the first pulses after a breakdown, there is a ramping period, where the field is briefly decreased to one fifth of the value before the BD and then it is asymptotically increased back to the target value during a course of \num{20} voltage steps with \num{100} pulses in each step as demonstrated in Figure \ref{fig:ramping}. The objective of the ramping is to reduce the possibility of cascades of secondary breakdowns.

\begin{figure}
\centering
\includegraphics[width=\linewidth]{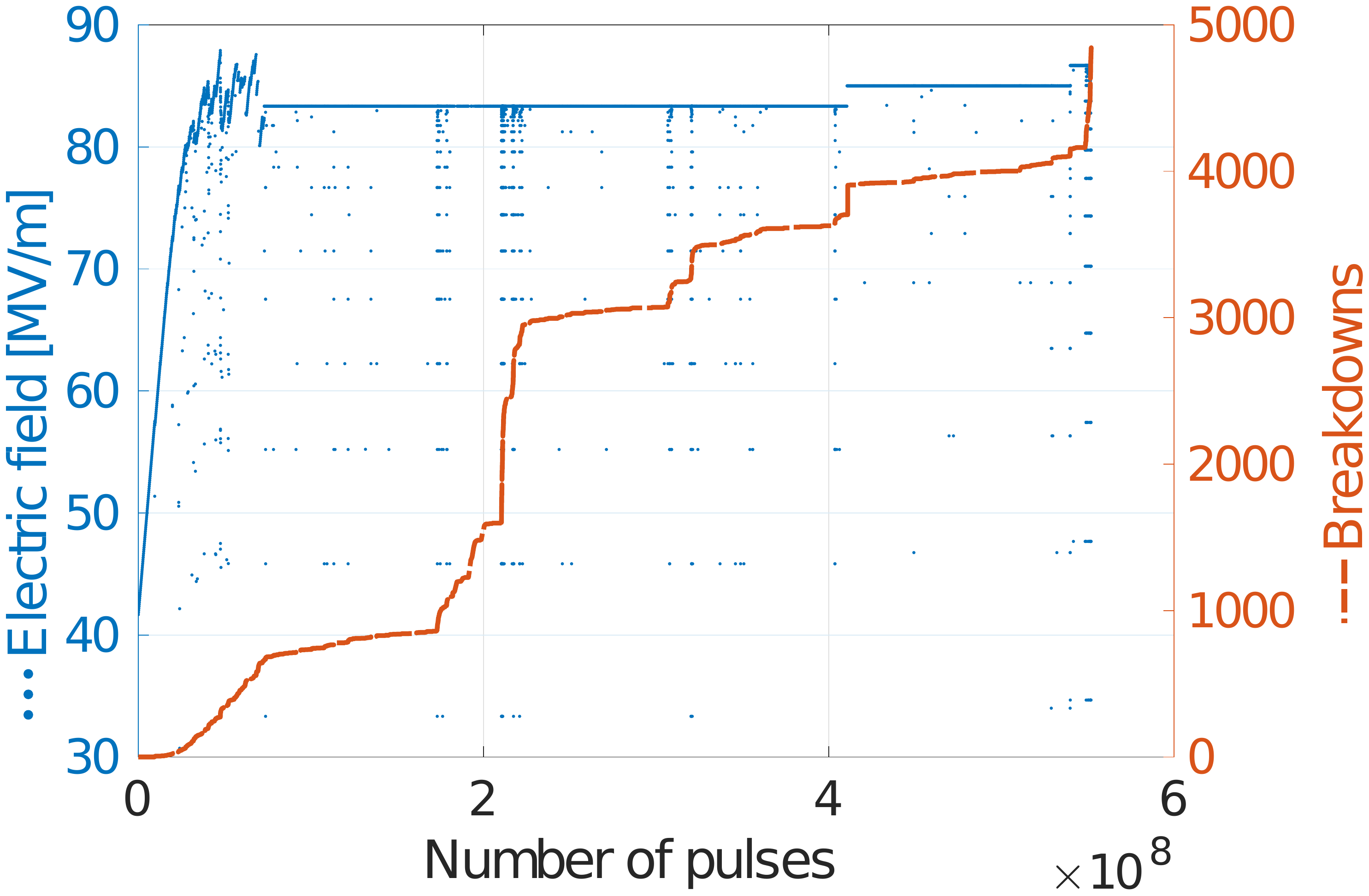}
\caption{\label{fig:soft_cu_history} Conditioning and first flat mode runs of Soft Cu Helsinki visualized. The graph shows evolution of the electric field and number of breakdowns as a function on pulses. The conditioning part, as well as the different flat mode runs are separated by red dashed lines.}
\end{figure}

After the electrode has reached a conditioned state -- i.e. the frequency of breakdowns is such that the algorithm keeps the electric field at a steady level, thus the field has saturated -- the measurements are continued in so-called flat mode runs.  In these runs the target voltage is set to a constant value where the breakdown rate is relatively stable. The different pulsing modes are visualized in Figure \ref{fig:soft_cu_history}.

\subsection{Electrodes}
The LES electrodes are of a cylindrical shape with the contact area of \SI{62}{\milli\meter} in diameter, while the diameter of the bottom of the electrodes, i.e. including the areas below the spacer, is \SI{80}{\milli\meter}. More technical detail on the shape of electrodes is available elsewhere \cite{Gudkov2014CDD:Disk}. Electrode thickness is greatest below the \SI{62}{\milli\meter} contact area, \SI{30}{\milli\meter}. The contact surface of the electrodes is diamond machined to have roughness below one micron, which is also the accuracy of the shoulder height which maintains the electrode separation via an aluminum oxide spacer.

As discussed previously, the material of interest is copper. The two main types of copper being tested are hard copper and soft copper. Hard copper is high-purity, oxygen-free electronic copper that is machined into the required shape. The grain diameter on the surface of hard copper is between \SIlist{10;100}{\micro\meter} (at least \num{4} on the ASTM E112 standard \cite{ASTMInternational2013ASTMSize}). Soft copper additionally undergoes a treatment first at \SI{1040}{\celsius} in hydrogen atmosphere and usually afterwards at \SI{650}{\celsius} in vacuum to breath out the hydrogen. However, the electrodes of the measurement Soft Cu CERN did not undergo the breath-out treatment. The average grain diameter of the Soft Cu CERN was estimated to be \SI[separate-uncertainty]{1.3 \pm 0.2}{\milli\meter} based on the Heyn Linear Intercept Procedure described in the ASTM E112 Standard.

\subsection{Breakdown localization}

The Pulsed DC System at CERN is additionally equipped with cameras for localization of the breakdowns. They are installed close to two viewports of the vacuum chamber, perpendicular to each other. Positions of the breakdowns are determined by the light emitted during each breakdown. If both cameras record a single line, the positioning of breakdown on the electrode surface can be determined. This technique allows to see real-time spatial distribution of breakdowns without the necessity of disassembling the vacuum chamber in order to see the BD spots which would cause several days' halt with the measurements. In these experiments, smaller electrodes with contact disk diameter of 40 mm were used. The localization algorithm is described in detail in Reference \cite{Profatilova2019BreakdownSystem}.
  
Collecting the data from the cameras, the high voltage generator, the oscilloscope and post-mortem microscopy, a wide range of parameters such as electric field, number of pulses between previous breakdown, distance between subsequent breakdowns and crater sizes on the anode and cathode surfaces can be analyzed for better understanding of each breakdown event. 

\begin{figure}[!htbp]
\centering
\includegraphics[width=\linewidth]{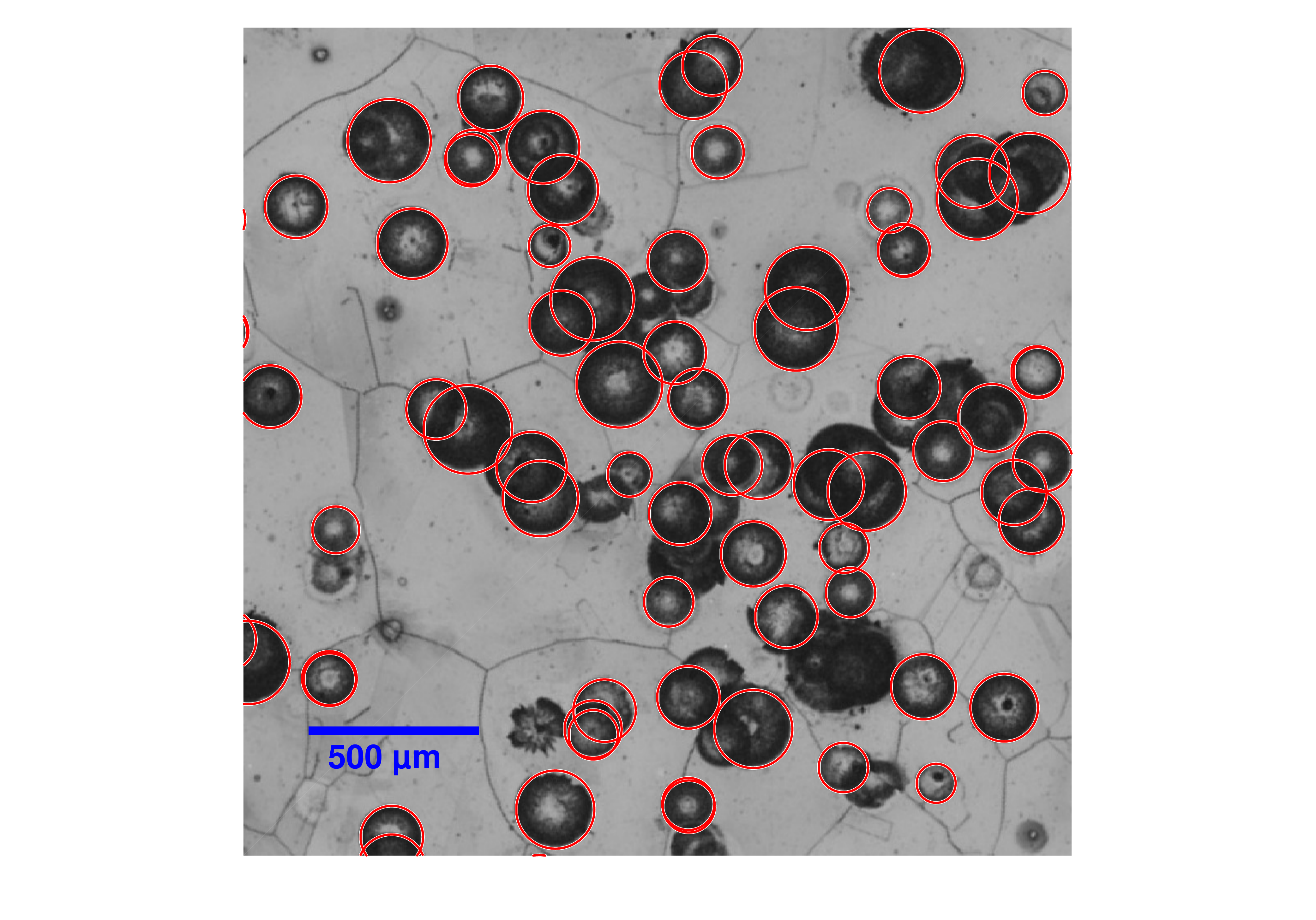}
\caption{\label{fig:BDseries} Optical microscope image of Soft Cu CERN surface with several BD craters visible and detected by machine vision.}
\end{figure}

Electrodes are also imaged both before breakdown experiments and post-mortem with optical microscopes. In these images, breakdowns can be seen as dark craters on the surface, though it is practically impossible to connect them to the pulse that caused the BD without the cameras for localization. Examples of breakdowns on a cathode surface of Soft Cu CERN are shown in Figure \ref{fig:BDseries}. Machine vision algorithm was used to detect the BD craters from the image and estimate their sizes and locations. The detection algorithm used two-stage circular Hough transform to identify the circular objects in an image. 

\subsection{Other parts of the measurements and analysis}
During the pulsing and breakdown experiments, also secondary parameters of the system were observed. These include monitoring the vacuum pressure as well as voltage and current waveforms. In some runs, we also used a mass spectrometer in order to detect residual particles in the vacuum.

The waveform and pulsing data are saved for each breakdown. This allows investigation of the breakdown current, voltage, timing within the pulse, short circuit width, pulse number and timing. Reference values are also saved for some non-breakdown pulses.

Also, simple simulations were used to understand the results. Particularly, the distribution of distances between breakdowns was simulated with Monte Carlo methods. That is, by repeatedly scattering series of \num{1000} BDs within \SI{3}{\milli\meter} from the edge of the electrode and calculating the distribution of distances between the subsequent spots in order to compare the distribution with the experimentally measured ones. The \SI{3}{\milli\meter} annulus was selected due to the fact that generally, especially for hard copper, more than \SI{90}{\percent} of the breakdowns lie within the \SI{3}{\milli\meter} from the edge. We will refer to these breakdowns as edge BDs. The increased BD density near the edges is currently explained by locally enhanced electric fields in the region \cite{Parker2002ElectricCapacitor,CatalanIzquierdo2017CapacitanceAnalysis}.

In addition, post-mortem analysis is performed to the surfaces for example to measure the electrode tilt and roughness with a profilometer or to investigate the BD craters with a Scanning White Light Interferometry microscope \cite{Kassamakov20173DInterferometry}.

A breakdown event concentrates tens millijoules of energy within a small area \cite{Latham1995HighPractice,Kyritsakis2018ThermalEmission}, which results in creation of the BD craters. It has been observed that a cathode crater, generated by a BD in LES with a \SI{60}{\micro\meter} gap, usually contains a pit that has a typical depth of \SI{1}{\micro\meter} and a radius of \SI{50}{\micro\meter}. At the edges of the pit, there is typically a \SI{50}{\micro\meter} thick annulus of molten and recrystallized material, which has numerous sharp edges and protrusions that serve as sites for additional BDs \cite{Wang1989RfStructures}.

\section{\label{sec:results}Results}
All the data presented below are the results of the flat mode runs with conditioned electrodes. The conditions of the experiment were kept constant as long as possible in order to collect enough data for statistical analysis. Most importantly, the pulsing voltage was kept constant except immediately after each breakdown, when the asymptotic ramping described earlier was used to ramp up the electric field from one fifth to the target voltage in \num{2000} pulses (\num{2100} for the setup in CERN).

\subsection{Pulses between breakdowns}
Number of pulses between two consecutive breakdowns were analyzed for four flat mode runs with different electrodes -- two sets in CERN and two in Helsinki. The key numbers of the runs are shown in Table \ref{tab:datasets}. Hard Cu CERN and Soft Cu CERN had a contact surface diameter of 40 mm in order to enable BD localization. For reference, the same analysis was also conducted for an RF run conducted with the X-band test stand at CERN \cite{Zennaro2017HighCLIC}.

\begin{table}
\centering
\caption{The four DC data sets and one RF run used for comparing the statistics on pulses between breakdowns.}
\label{tab:datasets}
\begin{tabular}{|l|c|c|c|c|}
\hline
\textbf{Name}       & \textbf{Pulse length} & \textbf{Electric field}       & \textbf{RepRate} \\ \hline 
Hard Cu CERN        & \SI{1}{\micro\second} & \SI{83}{\mega\volt/\meter}    & \SI{2000}{\hertz} \\ \hline 
Soft Cu CERN        & \SI{1}{\micro\second} & \SI{85}{\mega\volt/\meter}    & \SI{2000}{\hertz}\\ \hline 
Hard Cu Helsinki    & \SI{3}{\micro\second} & \SI{83}{\mega\volt/\meter}    & \SI{2000}{\hertz} \\ \hline 
Soft Cu Helsinki    & \SI{1}{\micro\second} & \SI{83}{\mega\volt/\meter}    & \SI{4000}{\hertz} \\ \hline 
RF Run CERN         & \SI{83}{\pico\second} & \SI{108}{\mega\volt/\meter}   & \SI{50}{\hertz} \\ \hhline{|=|=|=|=|} 
\textbf{Name}       & \textbf{BDs}          & \textbf{Pulses}               & \textbf{BDR} \\ \hline 
Hard Cu CERN        & \num{2 383}           & \num{3.15e7}                  & \num{7.55e-5} \\ \hline 
Soft Cu CERN        & \num{1 489}           & \num{2.03e7}                  & \num{7.34e-5} \\ \hline 
Hard Cu Helsinki    & \num{414}             & \num{9.69e7}                  & \num{4.27e-6} \\ \hline 
Soft Cu Helsinki    & \num{1 830}           & \num{1.98e8}                  & \num{9.24e-6} \\ \hline 
RF run CERN         & \num{93}              & \num{3.44e7}                  & \num{2.70e-6} \\ \hline  
\end{tabular}
\end{table}

\begin{figure}
\centering
\includegraphics[width=\linewidth]{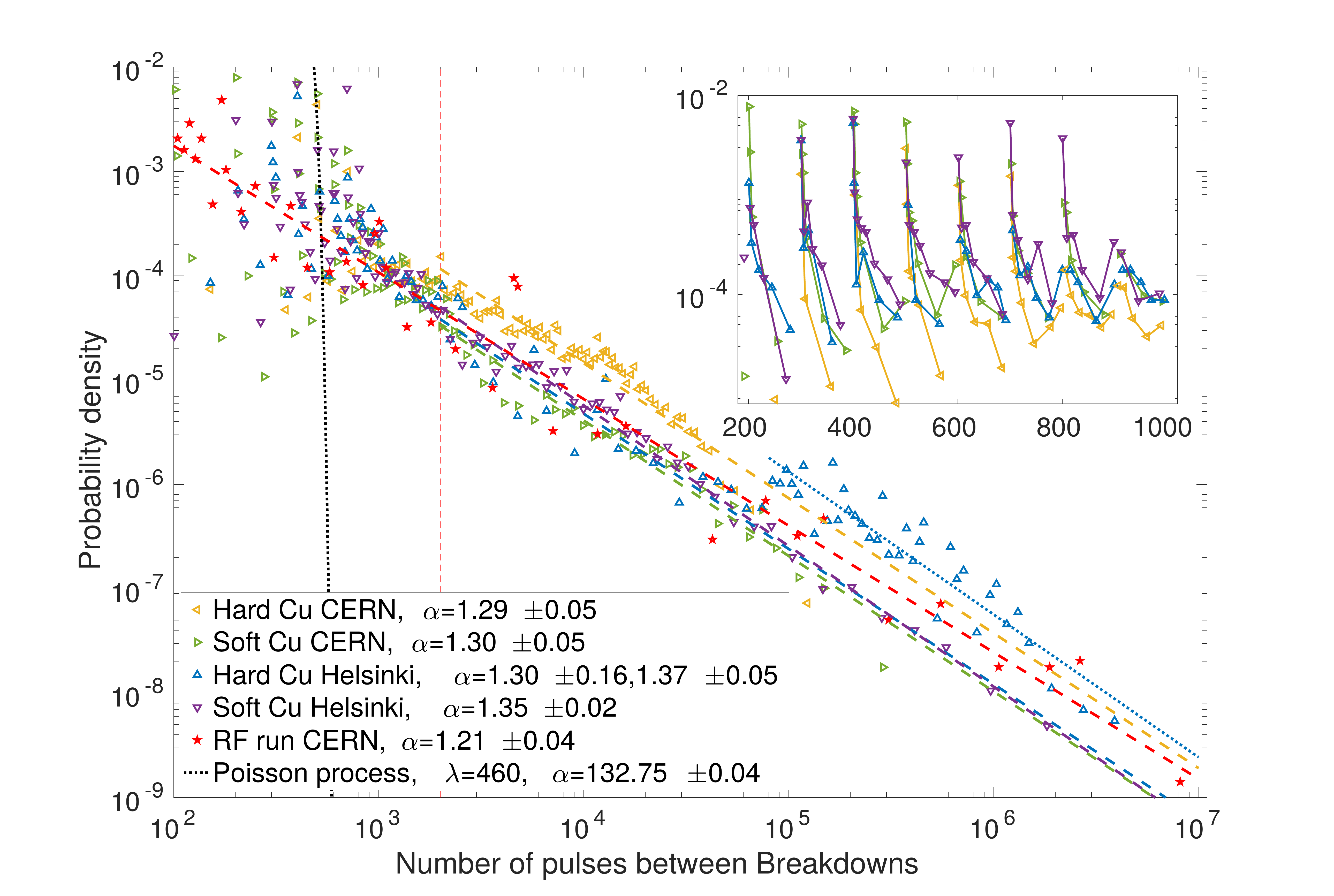}
\caption{\label{fig:pBetween_all} Probability distribution of pulses between breakdowns on doubly logarithmic scale. The black dotted line shows probability distribution function of a Poisson distribution for the comparison. A semilogarithmic inset shows a zoom-in of the distribution during the ramping when the voltage is between \SIlist{50;90}{\percent} (\numrange[range-phrase = --]{200}{1000} pulses) of the target voltage. The red dashed line indicates the end of the ramping phase. The uncertainties are obtained from the standard deviations of the slope values.}
\end{figure}

A probability distribution function of the number of pulses between two subsequent breakdowns was generated by collecting the events of a given number of pulses into logarithmically spaced bins in the pulse range. The results are shown in Figure \ref{fig:pBetween_all}. The graphs show that the probability for a breakdown to occur within the ramping period is remarkably higher than after the ramping, but the values vary significantly within this period, correlating with the ramping steps. Comparison shows that the points within the each step follow the linear part of Poisson distribution, except for the latest points of each step.

After the ramping has ended, the probability decreases linearly on the log--log scale. There are no big differences in the trends between the PDFs for different runs, except for the Hard Cu Helsinki, which has a jump in the probability at around $10^5$ pulses, but continues with almost the same slope even after the jump. The linear decay of the probabilities follow the power law $P(S) = kS^{-\alpha}$, with $\alpha \approx \num{1.30}$, which was similar for all the runs. The RF run does not have comparable steps in the ramping algorithm, though the results still nicely follow the power law with only slightly smaller slope.

The inset of Figure \ref{fig:pBetween_all} reveals that the probability of BDs shows a sawtooth-pattern with peaks at every 100 pulses -- exactly at the beginning of each ramping step. The sawtooth-pattern confirms an earlier qualitative observation that the breakdown probability increases whenever the pulsing is paused and the conditions are changed. In this case, the pause was a few seconds, which was required for the power supply to adjust to the new voltage for each ramping step. Due to the asymptotic ramping, the relative changes in the electric field are the largest in the first ramping steps (below \num{800} pulses), where the relative increase per step is more than \SI{3}{\percent}. Above \num{800} pulses in the ramping, the voltage is already above \SI{90}{\percent} of the target value. This explains why the clearest sawtooth-pattern is seen between \numlist{200;800} pulses, where both the relative change and the absolute voltage are large enough. It is important to note that by bypassing the ramping and starting directly at the target voltage, there would be even more breakdowns during the first pulses after the previous BD.

\subsection{Distances between breakdowns}

\begin{figure}
\centering
\includegraphics[width=\linewidth]{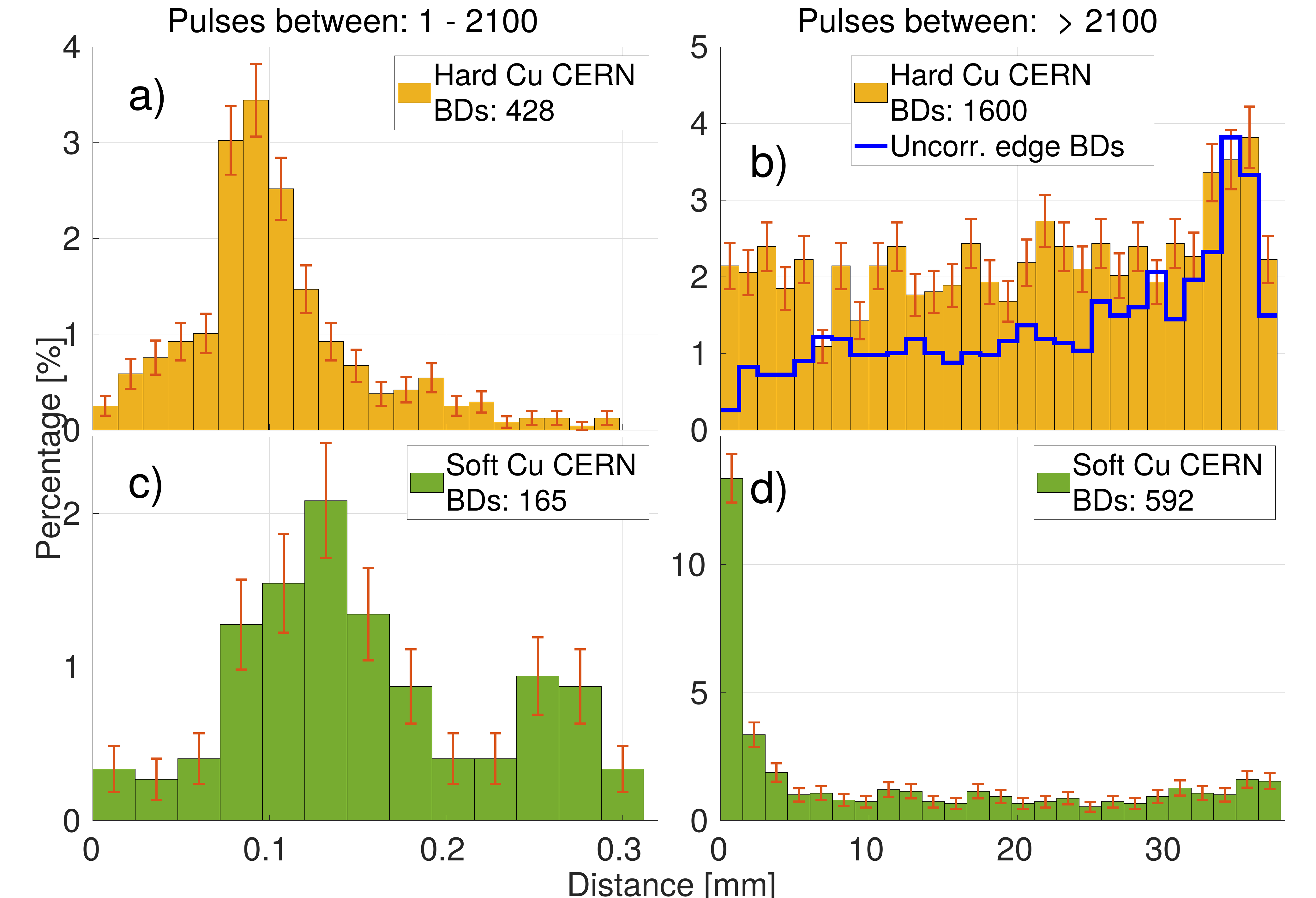}
\caption{\label{fig:alldistances} Distribution of distances between breakdowns for Hard and Soft Cu, grouped by pulses between breakdowns with error bars showing the standard deviation. a) and c) show the distributions for ramping pulses where the only features in the distributions are visible within the first \SI{300}{\micro\meter}. b) and d) show the distributions for all the BDs at all distances for BDs after the ramping period. In addition, b) shows the distribution distances between simulated uncorrelated edge BDs. The integral of each distribution equals to the fraction of breakdowns that fall within each pulse and distance range, so that both ranges combined equal \SI{100}{\percent}.}
\end{figure}

Distances between consecutive breakdowns were measured using the localization technique described in Section \ref{sec:exp}. Figure \ref{fig:alldistances} shows the probability for a breakdown to occur at a certain distance from the previous one (center to center) for Hard and Soft Cu. The left panel (Figures \ref{fig:alldistances}a and \ref{fig:alldistances}c) in the graph shows the distributions for breakdowns that occurred during ramping. They are mainly localized within the distance of \SI{300}{\micro\meter}, especially on Hard Cu surface, while almost no breakdowns are seen at the larger distances. The right panels (Figures \ref{fig:alldistances}b and \ref{fig:alldistances}d) show the distributions of distances between the breakdowns registered after the ramping mode was completed. There we see that the probabilities are almost equal for all the distances.
especially on Hard Cu, where the distribution is very similar to simulated, uncorrelated BD locations within \SI{3}{\milli\meter} from the edge of the electrode.

\begin{figure}[!htbp]
\centering
\includegraphics[width=\linewidth]{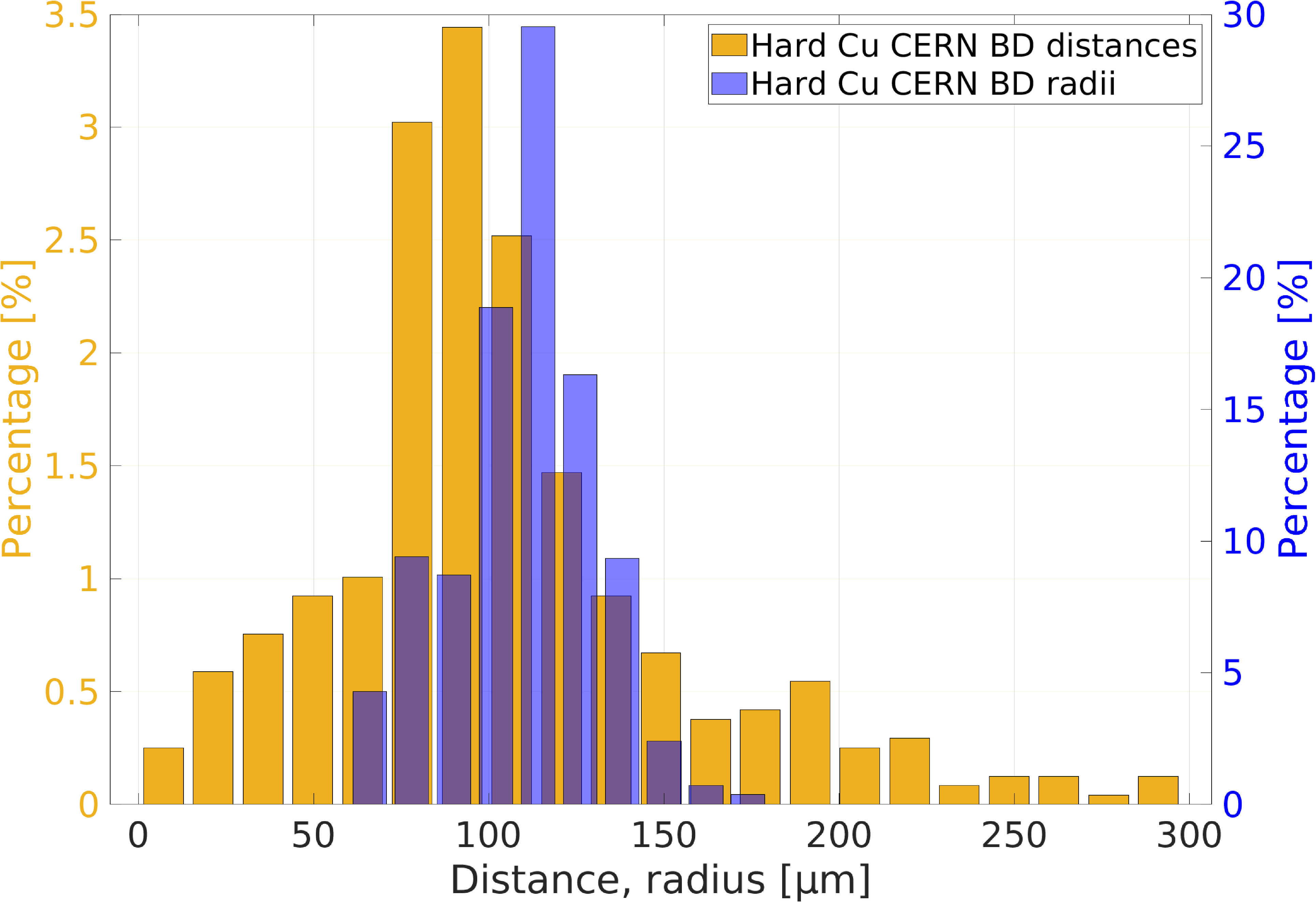}
\caption{\label{fig:distfits_soft} Distributions of distances between ramping breakdowns near the previous breakdown spot (same as Figure \ref{fig:alldistances}a)) and the radii of cathode breakdown craters compared, both from Hard Cu CERN.}
\end{figure}

In the subfigures \ref{fig:alldistances}a) and \ref{fig:alldistances}c), we see that there is an increased probability to have a breakdown at around \SI{100}{\micro\meter}, which happens to be close to the average radius of a BD crater on cathode, as seen in Figure. \ref{fig:distfits_soft}. This behaviour is very similar for both Hard and Soft Cu.

\begin{figure}
\centering
\includegraphics[width=\linewidth]{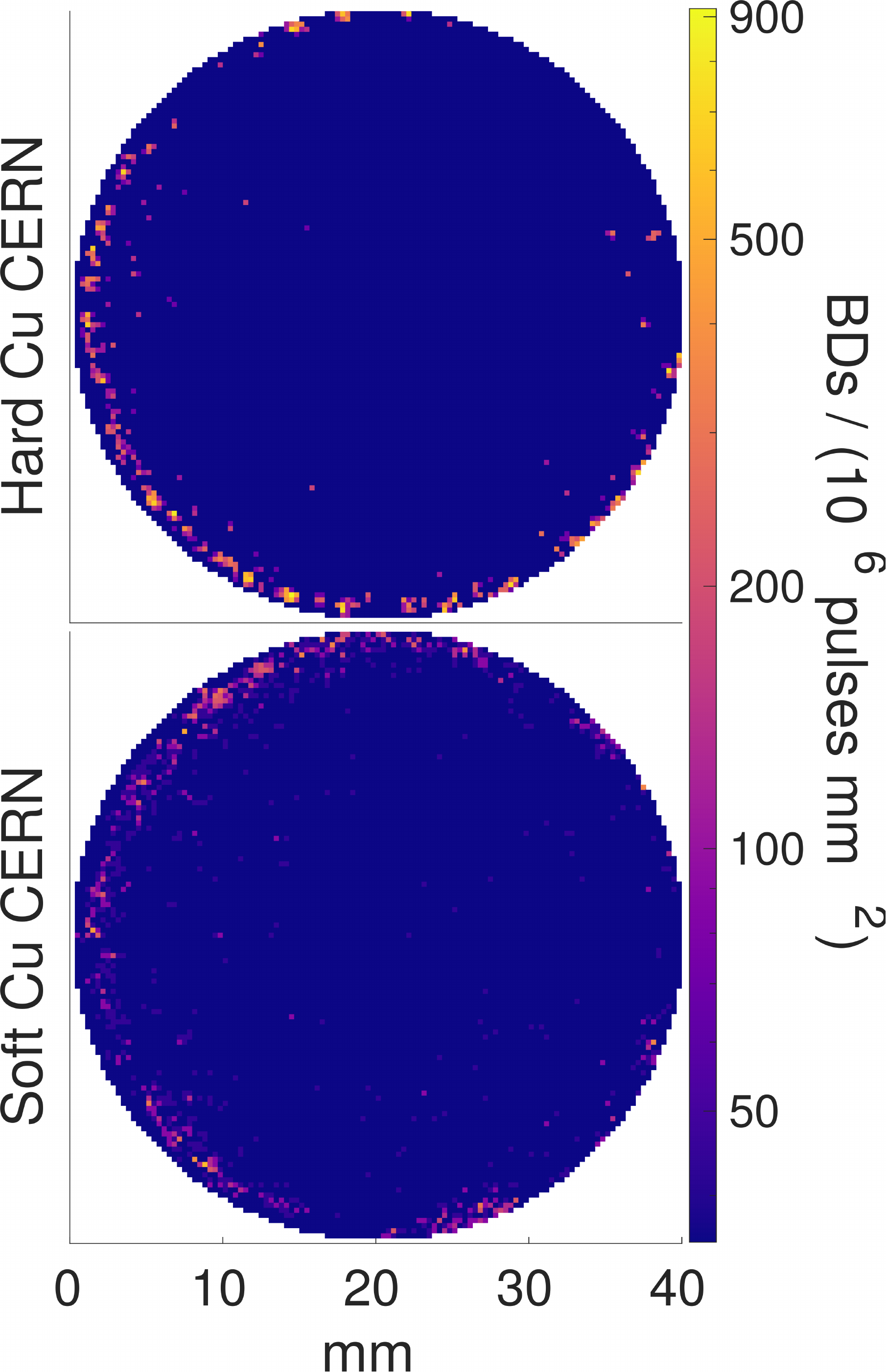}
\caption{\label{fig:heatplot} Density graph of the breakdown positions on Hard Cu CERN and Soft Cu CERN. The plots show that on Hard Cu, the BDs tend to appear in distinct sites, whereas on Soft Cu, the clusters are not as strong and the distribution is more widely spread (leading to lower BD density). It is also important to note that on both electrodes, majority of the BDs lie on within a few millimeters from the edge.}
\end{figure}

In the subfigures \ref{fig:alldistances}b) and \ref{fig:alldistances}d), however, we see large differences between the copper types. While the distribution for Hard Cu is more or less uniform, the Soft Cu still has an increased amount of BDs close to the previous one. At first, this seems to contradict the earlier observation reported in \cite{Korsback2019VacuumSystem}, which showed intense spatial clustering of BDs on Hard Cu, while on Soft Cu, the spatial BD distribution was much more uniform across the whole surface. However, a closer look at the locations of consecutive BDs reveals that in Soft Cu, it is common to have several consequent BDs within a close distance (less than a millimeter) from each other, after which the next breakdown can be anywhere. On Hard Cu, the BDs are more clustered, but it is rare to have more than one consecutive BD in the same cluster -- the next BD is more likely to occur in another cluster anywhere on the surface. The clustering effect is visualized in Figure \ref{fig:heatplot}, where we see that the BD density on Hard Cu exceeds \num{500} BDs per million pulses per \si{\milli\meter\squared} in several spots, whereas on Soft Cu there are only less prominent clusters and the BDs are more widespread outside of clusters.

In Figure \ref{fig:distfits_soft}, the distances between consecutive ramping breakdowns are shown again for the Hard Cu CERN. This time the distances are compared  with the size distribution of the cathode spots. The distributions show similar trends, with peaks around \SI{100}{\micro\meter}. Gaussian fits for the distributions yield medians of \SI[separate-uncertainty]{104 \pm 4}{\micro\meter} and \SI[separate-uncertainty]{110 \pm 1}{\micro\meter}, for the distances and radii, respectively.

\begin{figure}[!htbp]
\centering
\includegraphics[width=\linewidth]{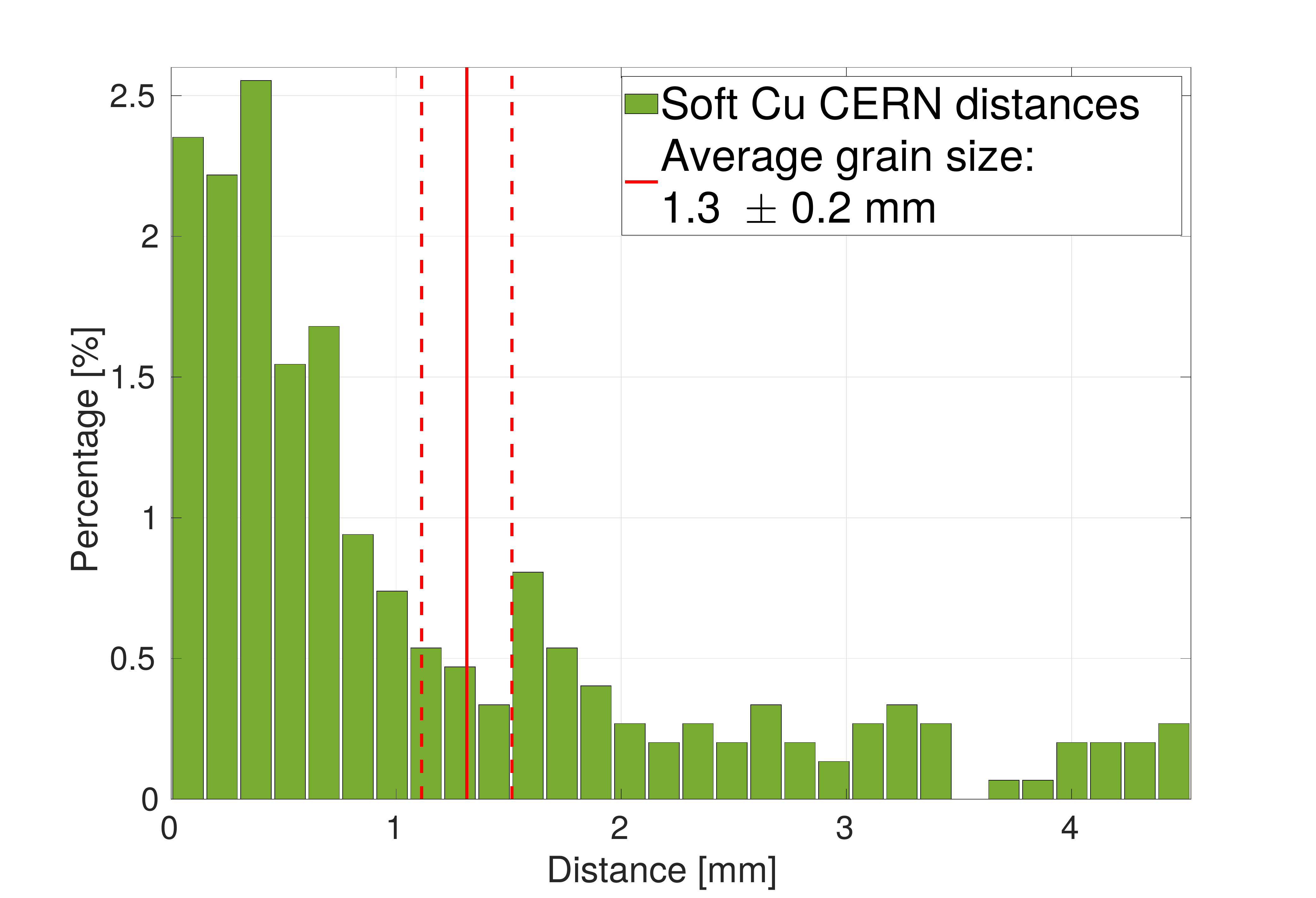}
\caption{\label{fig:soft_graindists} Distribution of distances between non-ramping breakdowns within \SI{4}{\milli\meter} from the previous breakdown spot on Soft Cu CERN, i.e. a zoom to the first three bars (distances \numrange[range-phrase = --]{0}{4.5}\,\si{\milli\meter}) of Figure \ref{fig:alldistances}d) compared with the average grain diameter distribution (solid line) and its standard deviation (dashed line).}
\end{figure}

Figure \ref{fig:soft_graindists} shows that a large part (\SI{32}{\percent}) of non-ramping breakdowns occur within the first \SI{1.5}{\milli\meter} from each other, after which the distribution is flat and close to zero for the rest of the \SI{38.5}{\milli\meter} of the surface. The \SI{1.5}{\milli\meter} mm cut-off value is close to the average grain diameter of \SI[separate-uncertainty]{1.3 \pm 0.2}{\milli\meter}

\section{Discussion}
The results show that the probability for a breakdown to occur is the highest within the next few hundred pulses after the previous one -- and within \SI{300}{\micro\meter} from the center of the previous BD spot. That is why the ramping has been introduced to keep the breakdown rate approximately stable. During the ramping period, the BD probability strongly follows the ramping steps so that it is the highest at the beginning of each step. The lowest probability within each ramping step is always the latest data point of the step. This value is higher when the electric field is higher at the corresponding ramping step. This is seen in overall ascending trend in the sawtooth function shown in the inset of Figure \ref{fig:pBetween_all}.

This, linked to the high localization of the consequent ramping BDs, shows that there is a correlation between the events. Since these breakdowns appear as follow-up events, we call them secondary breakdowns. The breakdowns, which take place after a large number of pulses and do not exhibit any spatial or temporal correlation with the preceding one, are called primary BD.

After the ramping, the BD probability decays linearly on the log--log scale, as a function of the number of pulses, with the slope being $\alpha\approx \num{1.33}$. The value is really close to the slope obtained with the previous system, reported in \cite{Wuensch2017StatisticsRegime} and also relatively close to the RF experiments in CERN. The observed jump at around 10$^5$ pulses in Hard Cu Helsinki plot is most probably an artifact from the changing of measurement period, which also happens every 10$^5$ pulses, granted that there was no BD.  This kind of power law behaviour is seen in various seemingly unrelated phenomena, such as avalanche size distribution and earthquake frequency -- and also the behaviour of dislocations in metals \cite{Papanikolaou2011UniversalityShape, Chrzan1994CriticalityCompounds}. Observed $\alpha$ in the references is typically \num{1.5}, so relatively close to the measured values. Important feature of this kind of behaviour is the universality across several magnitudes of scales.

These events that happen during the linear part of the Figure \ref{fig:pBetween_all} can be seen as primary breakdowns as they are mostly independent from one another and typically followed by secondary, highly dependent, BDs. Their initiation requires some local changes in the material, making that particular spot "hot". To understand the BD ignition, it is really important to understand what makes this particular spot more favourable for a BD than any other on the surface with an area larger than \SI{10}{\centi\meter\squared}. Earlier work have hypothesized linking this to dislocations piling up near the surface, causing formation of protrusions, which, in turn, enhance the electric field locally \cite{Engelberg2018StochasticFields}. The power law behaviour can be explained by dislocation avalanches.The relation between localized pre-breakdown field emission currents, hot spots and breakdown initiation was found in References \cite{Davies1973TheReview,Davies1966VacuumElectrodes}. The phenomenon of hot, breakdown-apt regions have also been observed in the cells of RF structures \cite{Adolphsen2001ProcessingNLCTA,DalForno2016RfStructures,Degiovanni2016ComparisonStructures} and been previously connected to surface contamination \cite{Mesyats2013EctonArc}.

\begin{figure}[!htbp]
\centering
\includegraphics[width=\linewidth]{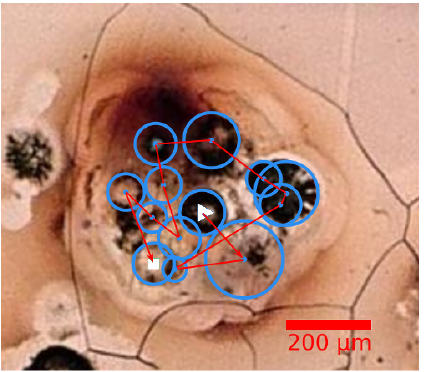}
\caption{\label{fig:BDseries48} Optical microscope image of a series of \num{13} consecutive soft copper BDs in a cluster on, bounded by a grain on a soft copper surface. $\triangleright$ and $\Box$ indicate the first and last BDs, respectively. The circle edges and centers were identified manually as the machine vision algorithm was not usable due to lack of contrast. The order of the BD sequence was obtained from the BD localization information.}
\end{figure}

The third main observation is the most probable distance between two consecutive breakdowns. If the BD would not fully destroy the underlying material, it would be natural that the next breakdown is likely to hit in the same crater again. Even more interesting is the finding of increased probability for a BD to occur at around \SI{100}{\micro\meter} from the previous spot. This happens to be very close to the average radius of a breakdown crater. This suggests that the molten areas near the edges of the crater form ideal conditions for next breakdowns to happen as already observed in \cite{Wang1989RfStructures}. The phenomenon can actually be seen in the post-mortem surface images, showing chains of breakdowns exactly one crater radius apart from each other as seen in Figures \ref{fig:BDseries} and \ref{fig:BDseries48}. The small difference in the mean values could be explained by systematic uncertainties in the BD crater recognition algorithm and its definition of BD crater "edge".

\begin{figure}[!htbp]
\centering
\includegraphics[width=\linewidth]{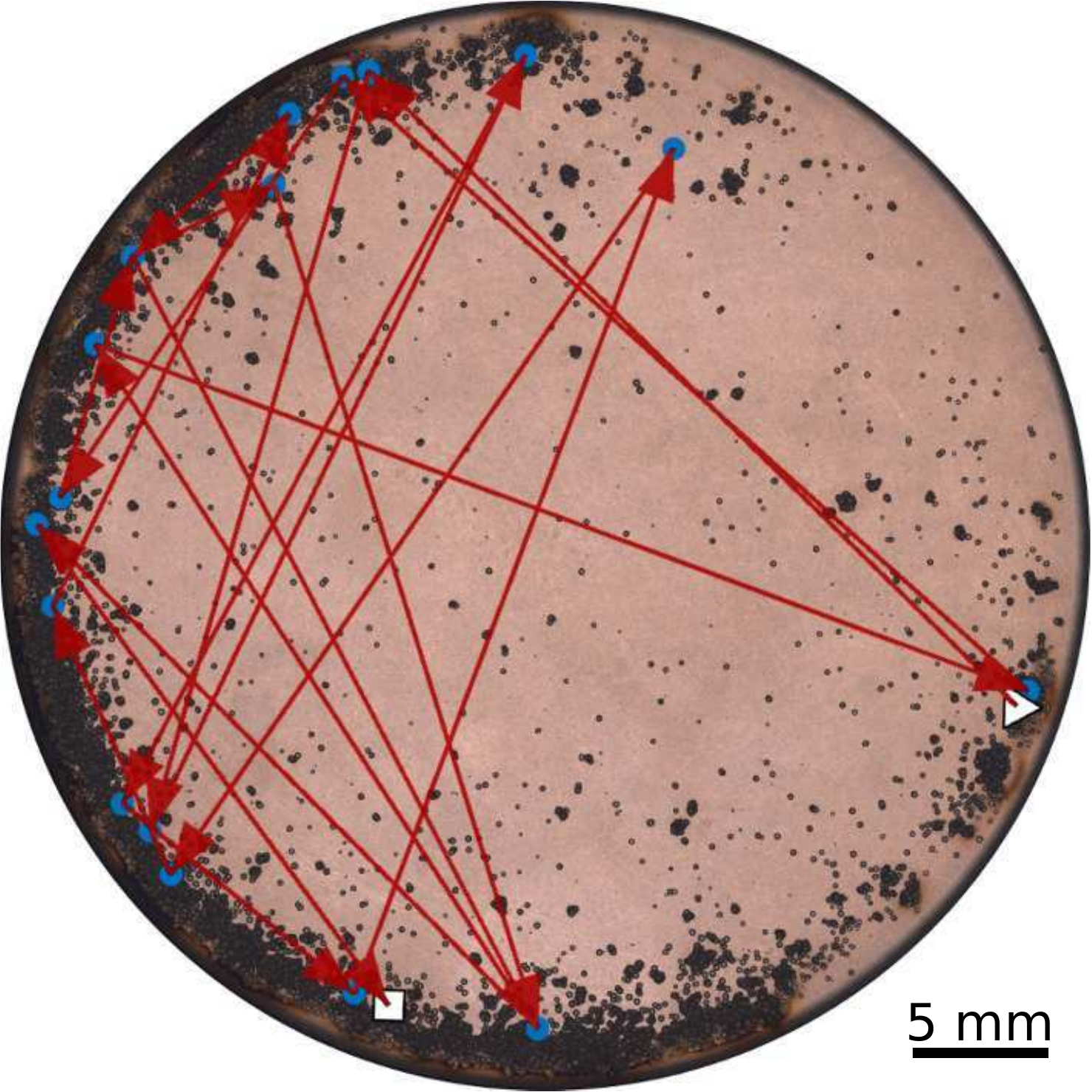}
\caption{\label{fig:BDseriesHard} Optical microscope image of the full surface of the Hard Cu CERN cathode. On the surface, a series of \num{19} consecutive BDs is shown, each in a separate site randomly distributed near the edge of the cathode. $\triangleright$ and $\Box$ indicate the first and last BDs, respectively.}
\end{figure}

The subfigures \ref{fig:alldistances}b) and \ref{fig:alldistances}d) show clear differences between Hard and Soft Cu. The Hard distribution pretty much follows the shape of random, uncorrelated edge BDs, as the simulation result shows in the same figure, suggesting that BDs on hard Cu surfaces after the ramping period are mostly uncorrelated, stochastic events. Soft Cu BDs however, show correlation even after the ramping period, as most of the BDs fall within a few millimeters from the previous one. This can also be observed when studying series of consecutive non-ramping BDs that are located within \SI{1}{\milli\meter} from each other. In Hard Cu, less than \SI{2}{\percent} of the BDs fall into this category, while in Soft Cu, the fraction is around \SI{10}{\percent} of the events (counting the series with at least two BDs).

Also previous research suggests that the breakdowns can be classified as either primary or secondary events \cite{Wuensch2017StatisticsRegime,Mesyats2013EctonArc}. These are defined based on the physical distance and pulses between the events. The primary BDs initiate when some features, such as dislocations congest near surface in a particular hot grain. With hard copper, the grain diameter is in the order of \SI{10}{\micro\meter}, so one breakdown does diminish the whole area of the grain and it becomes inactive for a large number of pulses. This can be understood in terms of dislocation activities. A new set of dislocations needs to be activated before the grain becomes active again. Thus, it takes a large number of pulses before that spot can recover and a new primary breakdown can occur there. During this recovery time, it is more probable that the dislocations are mobilized in some other grain, leading to a breakdown in this spot. Figure \ref{fig:BDseriesHard} shows an example of this kind of a series of \num{19} consecutive primary BDs randomly scattered on the electrode surface (though nearly all of them are near the edge).

With soft copper, the grain diameter can be up to a few millimeters, allowing multiple primary breakdowns to occur close to each other before the grain is quenched. Secondary breakdowns are those that occur within the same breakdown crater, also typically close in terms of pulses in between. Figure \ref{fig:BDseries48} shows an example of a series of \num{13} consecutive BDs (including some ramping BDs) within one grain.

The different behaviour between surfaces with different grain sizes fits the dislocation hypothesis suggested in References \cite{Nordlund2012DefectFields,Engelberg2018StochasticFields}, which is linked to the grain size: the dislocations are known to get pinned or annihilated at the grain boundaries \cite{Hull2011IntroductionDislocations}. With hard copper, presumably only some of the grains have mobile dislocations at a time, and once a BD has occurred on one, the source of protrusions -- a hot spot -- is at least temporarily deactivated. 

The small size of grains allows for fewer mobile dislocations, since they have higher probability to be stopped (pinned) at the grain boundaries. The large grains of well-annealed Cu offer freer movement of dislocations, which can arrive at the surface participating, for instance, in growth of surface asperities \cite{Pohjonen2012AnalyticalStress}.

With soft copper again, most grains are so large that they are prone to having at least some mobile dislocations. When a BD occurs in this kind of a hot grain, one event is not enough to quench the whole surface of the grain and some other BDs are likely to occur due to nearby dislocations. The dislocation hypothesis is also supported by the temperature dependence of copper breakdown susceptibility \cite{Cahill2018HighCavities}.

\section{Conclusions}
Vacuum arc breakdowns between two Cu electrodes were generated by DC pulses in order to understand their generation processes and how to limit the frequency of the BD events. In the analysis, the pulses between breakdowns, breakdown locations and their correlation to cathode crater size were compared over various flat mode measurement runs and two Cu types, both at the Accelerator Laboratory of the University of Helsinki and in CERN.

The results support the previous observation of independent \textit{primary} and correlated \textit{secondary} follow-up breakdowns. The BD locations show differences between hard and soft copper, which can be explained by grain size difference and dislocations. The dislocation hypothesis is also supported by the power law trend on pulses between breakdowns of the primary events.

For future work, it will be important to analytically understand the differences in the behaviour between Hard Cu and Soft Cu and to investigate the ramping algorithm so that it minimizes the breakdown rate without affecting the pulsing efficiency too much. Also, understanding the contaminants on electrode surfaces can play important role in understanding the full span of the breakdown processes \cite{Abe2018DirectCavity}.

\section{Acknowledgements}

The research was funded by the K-contract between Helsinki Institute of Physics and CERN. The optical microscopy was performed by Timo Hilden and Essi Kangasharju at the Detector Laboratory of the Helsinki Institute of Physics, as well as by Enrique Rodriguez Castro	at CERN. The electrode surface profilometry measurements were performed by Jouni Heino, also at the Detector Laboratory of the Helsinki Institute of Physics. The breakdown craters were also studied with the Scanning White Light Microscope by Anton Nolvi and Ivan Kassamakov at the Electronics Research Laboratory of the University of Helsinki.

\clearpage

\bibliography{_main.bib}

\begin{thebibliography}{48}%
\makeatletter
\providecommand \@ifxundefined [1]{%
 \@ifx{#1\undefined}
}%
\providecommand \@ifnum [1]{%
 \ifnum #1\expandafter \@firstoftwo
 \else \expandafter \@secondoftwo
 \fi
}%
\providecommand \@ifx [1]{%
 \ifx #1\expandafter \@firstoftwo
 \else \expandafter \@secondoftwo
 \fi
}%
\providecommand \natexlab [1]{#1}%
\providecommand \enquote  [1]{``#1''}%
\providecommand \bibnamefont  [1]{#1}%
\providecommand \bibfnamefont [1]{#1}%
\providecommand \citenamefont [1]{#1}%
\providecommand \href@noop [0]{\@secondoftwo}%
\providecommand \href [0]{\begingroup \@sanitize@url \@href}%
\providecommand \@href[1]{\@@startlink{#1}\@@href}%
\providecommand \@@href[1]{\endgroup#1\@@endlink}%
\providecommand \@sanitize@url [0]{\catcode `\\12\catcode `\$12\catcode
  `\&12\catcode `\#12\catcode `\^12\catcode `\_12\catcode `\%12\relax}%
\providecommand \@@startlink[1]{}%
\providecommand \@@endlink[0]{}%
\providecommand \url  [0]{\begingroup\@sanitize@url \@url }%
\providecommand \@url [1]{\endgroup\@href {#1}{\urlprefix }}%
\providecommand \urlprefix  [0]{URL }%
\providecommand \Eprint [0]{\href }%
\providecommand \doibase [0]{http://dx.doi.org/}%
\providecommand \selectlanguage [0]{\@gobble}%
\providecommand \bibinfo  [0]{\@secondoftwo}%
\providecommand \bibfield  [0]{\@secondoftwo}%
\providecommand \translation [1]{[#1]}%
\providecommand \BibitemOpen [0]{}%
\providecommand \bibitemStop [0]{}%
\providecommand \bibitemNoStop [0]{.\EOS\space}%
\providecommand \EOS [0]{\spacefactor3000\relax}%
\providecommand \BibitemShut  [1]{\csname bibitem#1\endcsname}%
\let\auto@bib@innerbib\@empty
\bibitem [{\citenamefont {Boxman}\ \emph {et~al.}(1996)\citenamefont {Boxman},
  \citenamefont {Sanders},\ and\ \citenamefont
  {Martin}}]{Boxman1996HandbookApplications}%
  \BibitemOpen
  \bibfield  {author} {\bibinfo {author} {\bibfnamefont {Raymond~L}\
  \bibnamefont {Boxman}}, \bibinfo {author} {\bibfnamefont {David~M}\
  \bibnamefont {Sanders}}, \ and\ \bibinfo {author} {\bibfnamefont {Philip~J}\
  \bibnamefont {Martin}},\ }\href@noop {} {\emph {\bibinfo {title} {{Handbook
  of vacuum arc science {\&} technology: fundamentals and applications}}}}\
  (\bibinfo  {publisher} {William Andrew},\ \bibinfo {year} {1996})\BibitemShut
  {NoStop}%
\bibitem [{\citenamefont {Latham}(1995)}]{Latham1995HighPractice}%
  \BibitemOpen
  \bibfield  {author} {\bibinfo {author} {\bibfnamefont {R.~V.}\ \bibnamefont
  {Latham}},\ }\href@noop {} {\emph {\bibinfo {title} {{High voltage vacuum
  insulation: Basic concepts and technological practice}}}}\ (\bibinfo
  {publisher} {Elsevier},\ \bibinfo {year} {1995})\BibitemShut {NoStop}%
\bibitem [{\citenamefont {Falkingham}\ and\ \citenamefont
  {Montillet}(2004)}]{Falkingham2004AConnection}%
  \BibitemOpen
  \bibfield  {author} {\bibinfo {author} {\bibfnamefont {L~T}\ \bibnamefont
  {Falkingham}}\ and\ \bibinfo {author} {\bibfnamefont {G~F}\ \bibnamefont
  {Montillet}},\ }\bibfield  {title} {\enquote {\bibinfo {title} {{A history of
  fifty years of vacuum interrupter development-(the English connection)}},}\
  }in\ \href@noop {} {\emph {\bibinfo {booktitle} {IEEE Power Engineering
  Society General Meeting, 2004.}}}\ (\bibinfo {year} {2004})\ pp.\ \bibinfo
  {pages} {706--711}\BibitemShut {NoStop}%
\bibitem [{\citenamefont {McCracken}(1980)}]{McCracken1980ATokamaks}%
  \BibitemOpen
  \bibfield  {author} {\bibinfo {author} {\bibfnamefont {G~M}\ \bibnamefont
  {McCracken}},\ }\bibfield  {title} {\enquote {\bibinfo {title} {{A review of
  the experimental evidence for arcing and sputtering in tokamaks}},}\
  }\href@noop {} {\bibfield  {journal} {\bibinfo  {journal} {Journal of Nuclear
  Materials}\ }\textbf {\bibinfo {volume} {93}},\ \bibinfo {pages} {3--16}
  (\bibinfo {year} {1980})}\BibitemShut {NoStop}%
\bibitem [{\citenamefont {Dyke}\ \emph {et~al.}(1953)\citenamefont {Dyke},
  \citenamefont {Trolan}, \citenamefont {Martin},\ and\ \citenamefont
  {Barbour}}]{Dyke1953TheInitiation}%
  \BibitemOpen
  \bibfield  {author} {\bibinfo {author} {\bibfnamefont {Walter~P}\
  \bibnamefont {Dyke}}, \bibinfo {author} {\bibfnamefont {J~K}\ \bibnamefont
  {Trolan}}, \bibinfo {author} {\bibfnamefont {E~E}\ \bibnamefont {Martin}}, \
  and\ \bibinfo {author} {\bibfnamefont {J~P}\ \bibnamefont {Barbour}},\
  }\bibfield  {title} {\enquote {\bibinfo {title} {{The field emission
  initiated vacuum arc. I. Experiments on arc initiation}},}\ }\href@noop {}
  {\bibfield  {journal} {\bibinfo  {journal} {Physical review}\ }\textbf
  {\bibinfo {volume} {91}},\ \bibinfo {pages} {1043} (\bibinfo {year}
  {1953})}\BibitemShut {NoStop}%
\bibitem [{\citenamefont {Charbonnier}\ \emph {et~al.}(1967)\citenamefont
  {Charbonnier}, \citenamefont {Bennette},\ and\ \citenamefont
  {Swanson}}]{Charbonnier1967ElectricalTheory}%
  \BibitemOpen
  \bibfield  {author} {\bibinfo {author} {\bibfnamefont {Francis~M}\
  \bibnamefont {Charbonnier}}, \bibinfo {author} {\bibfnamefont {Carol~J}\
  \bibnamefont {Bennette}}, \ and\ \bibinfo {author} {\bibfnamefont
  {Lynwood~W}\ \bibnamefont {Swanson}},\ }\bibfield  {title} {\enquote
  {\bibinfo {title} {{Electrical breakdown between metal electrodes in high
  vacuum. I. Theory}},}\ }\href@noop {} {\bibfield  {journal} {\bibinfo
  {journal} {Journal of Applied Physics}\ }\textbf {\bibinfo {volume} {38}},\
  \bibinfo {pages} {627--633} (\bibinfo {year} {1967})}\BibitemShut {NoStop}%
\bibitem [{\citenamefont {Schade}\ and\ \citenamefont
  {Shmelev}(2003)}]{Schade2003NumericalField}%
  \BibitemOpen
  \bibfield  {author} {\bibinfo {author} {\bibfnamefont {Ekkehard}\
  \bibnamefont {Schade}}\ and\ \bibinfo {author} {\bibfnamefont
  {Dmytry~Leonidovich}\ \bibnamefont {Shmelev}},\ }\bibfield  {title} {\enquote
  {\bibinfo {title} {{Numerical simulation of high-current vacuum arcs with an
  external axial magnetic field}},}\ }\href@noop {} {\bibfield  {journal}
  {\bibinfo  {journal} {IEEE Transactions on Plasma Science}\ }\textbf
  {\bibinfo {volume} {31}},\ \bibinfo {pages} {890--901} (\bibinfo {year}
  {2003})}\BibitemShut {NoStop}%
\bibitem [{\citenamefont {Zhou}\ \emph {et~al.}(2019)\citenamefont {Zhou},
  \citenamefont {Kyritsakis}, \citenamefont {Wang}, \citenamefont {Li},
  \citenamefont {Geng},\ and\ \citenamefont
  {Djurabekova}}]{Zhou2019DirectResolution}%
  \BibitemOpen
  \bibfield  {author} {\bibinfo {author} {\bibfnamefont {Zhipeng}\ \bibnamefont
  {Zhou}}, \bibinfo {author} {\bibfnamefont {Andreas}\ \bibnamefont
  {Kyritsakis}}, \bibinfo {author} {\bibfnamefont {Zhenxing}\ \bibnamefont
  {Wang}}, \bibinfo {author} {\bibfnamefont {Yi}~\bibnamefont {Li}}, \bibinfo
  {author} {\bibfnamefont {Yingsan}\ \bibnamefont {Geng}}, \ and\ \bibinfo
  {author} {\bibfnamefont {Flyura}\ \bibnamefont {Djurabekova}},\ }\bibfield
  {title} {\enquote {\bibinfo {title} {{Direct observation of vacuum arc
  evolution with nanosecond resolution}},}\ }\href {\doibase
  10.1038/s41598-019-44191-6} {\  (\bibinfo {year} {2019}),\
  10.1038/s41598-019-44191-6}\BibitemShut {NoStop}%
\bibitem [{\citenamefont {Barengolts}\ \emph {et~al.}(2018)\citenamefont
  {Barengolts}, \citenamefont {Mesyats}, \citenamefont {Oreshkin},
  \citenamefont {Oreshkin}, \citenamefont {Khishchenko}, \citenamefont
  {Uimanov},\ and\ \citenamefont
  {Tsventoukh}}]{Barengolts2018MechanismStructures}%
  \BibitemOpen
  \bibfield  {author} {\bibinfo {author} {\bibfnamefont {S.~A.}\ \bibnamefont
  {Barengolts}}, \bibinfo {author} {\bibfnamefont {V.~G.}\ \bibnamefont
  {Mesyats}}, \bibinfo {author} {\bibfnamefont {V.~I.}\ \bibnamefont
  {Oreshkin}}, \bibinfo {author} {\bibfnamefont {E.~V.}\ \bibnamefont
  {Oreshkin}}, \bibinfo {author} {\bibfnamefont {K.~V.}\ \bibnamefont
  {Khishchenko}}, \bibinfo {author} {\bibfnamefont {I.~V.}\ \bibnamefont
  {Uimanov}}, \ and\ \bibinfo {author} {\bibfnamefont {M.~M.}\ \bibnamefont
  {Tsventoukh}},\ }\bibfield  {title} {\enquote {\bibinfo {title} {{Mechanism
  of vacuum breakdown in radio-frequency accelerating structures}},}\ }\href
  {\doibase 10.1103/PhysRevAccelBeams.21.061004} {\bibfield  {journal}
  {\bibinfo  {journal} {Physical Review Accelerators and Beams}\ } (\bibinfo
  {year} {2018}),\ 10.1103/PhysRevAccelBeams.21.061004}\BibitemShut {NoStop}%
\bibitem [{\citenamefont {Kyritsakis}\ \emph {et~al.}(2018)\citenamefont
  {Kyritsakis}, \citenamefont {Veske}, \citenamefont {Eimre}, \citenamefont
  {Zadin},\ and\ \citenamefont {Djurabekova}}]{Kyritsakis2018ThermalEmission}%
  \BibitemOpen
  \bibfield  {author} {\bibinfo {author} {\bibfnamefont {Andreas}\ \bibnamefont
  {Kyritsakis}}, \bibinfo {author} {\bibfnamefont {Mihkel}\ \bibnamefont
  {Veske}}, \bibinfo {author} {\bibfnamefont {Kristjan}\ \bibnamefont {Eimre}},
  \bibinfo {author} {\bibfnamefont {Vahur}\ \bibnamefont {Zadin}}, \ and\
  \bibinfo {author} {\bibfnamefont {Flyura}\ \bibnamefont {Djurabekova}},\
  }\bibfield  {title} {\enquote {\bibinfo {title} {{Thermal runaway of metal
  nano-tips during intense electron emission}},}\ }\href@noop {} {\bibfield
  {journal} {\bibinfo  {journal} {Journal of Physics D: Applied Physics}\
  }\textbf {\bibinfo {volume} {51}},\ \bibinfo {pages} {225203} (\bibinfo
  {year} {2018})}\BibitemShut {NoStop}%
\bibitem [{\citenamefont {Nordlund}\ and\ \citenamefont
  {Djurabekova}(2012)}]{Nordlund2012DefectFields}%
  \BibitemOpen
  \bibfield  {author} {\bibinfo {author} {\bibfnamefont {Kai}\ \bibnamefont
  {Nordlund}}\ and\ \bibinfo {author} {\bibfnamefont {Flyura}\ \bibnamefont
  {Djurabekova}},\ }\bibfield  {title} {\enquote {\bibinfo {title} {{Defect
  model for the dependence of breakdown rate on external electric fields}},}\
  }\href@noop {} {\bibfield  {journal} {\bibinfo  {journal} {Physical Review
  Special Topics-Accelerators and Beams}\ }\textbf {\bibinfo {volume} {15}},\
  \bibinfo {pages} {71002} (\bibinfo {year} {2012})}\BibitemShut {NoStop}%
\bibitem [{\citenamefont {Engelberg}\ \emph {et~al.}(2018)\citenamefont
  {Engelberg}, \citenamefont {Ashkenazy},\ and\ \citenamefont
  {Assaf}}]{Engelberg2018StochasticFields}%
  \BibitemOpen
  \bibfield  {author} {\bibinfo {author} {\bibfnamefont {Eliyahu~Zvi}\
  \bibnamefont {Engelberg}}, \bibinfo {author} {\bibfnamefont {Yinon}\
  \bibnamefont {Ashkenazy}}, \ and\ \bibinfo {author} {\bibfnamefont {Michael}\
  \bibnamefont {Assaf}},\ }\bibfield  {title} {\enquote {\bibinfo {title}
  {{Stochastic model of breakdown nucleation under intense electric fields}},}\
  }\href@noop {} {\bibfield  {journal} {\bibinfo  {journal} {Physical review
  letters}\ }\textbf {\bibinfo {volume} {120}},\ \bibinfo {pages} {124801}
  (\bibinfo {year} {2018})}\BibitemShut {NoStop}%
\bibitem [{\citenamefont {Burrows}\ \emph {et~al.}(2018)\citenamefont
  {Burrows}, \citenamefont {Catalan-Lasheras}, \citenamefont {Linssen},
  \citenamefont {Petric}, \citenamefont {Robson}, \citenamefont {Schulte},
  \citenamefont {Sicking},\ and\ \citenamefont
  {Stapnes}}]{Burrows2018TheReport}%
  \BibitemOpen
  \bibfield  {author} {\bibinfo {author} {\bibfnamefont {P}~\bibnamefont
  {Burrows}}, \bibinfo {author} {\bibfnamefont {Nuria}\ \bibnamefont
  {Catalan-Lasheras}}, \bibinfo {author} {\bibfnamefont {Lucie}\ \bibnamefont
  {Linssen}}, \bibinfo {author} {\bibfnamefont {M}~\bibnamefont {Petric}},
  \bibinfo {author} {\bibfnamefont {Aidan}\ \bibnamefont {Robson}}, \bibinfo
  {author} {\bibfnamefont {Daniel}\ \bibnamefont {Schulte}}, \bibinfo {author}
  {\bibfnamefont {Eva}\ \bibnamefont {Sicking}}, \ and\ \bibinfo {author}
  {\bibfnamefont {Steinar}\ \bibnamefont {Stapnes}},\ }\bibfield  {title}
  {\enquote {\bibinfo {title} {{The Compact Linear e+ e- Collider (CLIC) 2018
  Summary Report}},}\ }\href@noop {} {\bibfield  {journal} {\bibinfo  {journal}
  {CERN Yellow Reports: Monographs}\ } (\bibinfo {year} {2018})}\BibitemShut
  {NoStop}%
\bibitem [{\citenamefont {Aicheler}\ \emph {et~al.}(2012)\citenamefont
  {Aicheler}, \citenamefont {Burrows}, \citenamefont {Draper}, \citenamefont
  {Garvey}, \citenamefont {Lebrun}, \citenamefont {Peach}, \citenamefont
  {Phinney}, \citenamefont {Schmickler}, \citenamefont {Schulte},\ and\
  \citenamefont {Toge}}]{Aicheler2012AReport}%
  \BibitemOpen
  \bibfield  {author} {\bibinfo {author} {\bibfnamefont {Markus}\ \bibnamefont
  {Aicheler}}, \bibinfo {author} {\bibfnamefont {P}~\bibnamefont {Burrows}},
  \bibinfo {author} {\bibfnamefont {M}~\bibnamefont {Draper}}, \bibinfo
  {author} {\bibfnamefont {Terence}\ \bibnamefont {Garvey}}, \bibinfo {author}
  {\bibfnamefont {P}~\bibnamefont {Lebrun}}, \bibinfo {author} {\bibfnamefont
  {K}~\bibnamefont {Peach}}, \bibinfo {author} {\bibfnamefont {N}~\bibnamefont
  {Phinney}}, \bibinfo {author} {\bibfnamefont {H}~\bibnamefont {Schmickler}},
  \bibinfo {author} {\bibfnamefont {Daniel}\ \bibnamefont {Schulte}}, \ and\
  \bibinfo {author} {\bibfnamefont {N}~\bibnamefont {Toge}},\ }\href {\doibase
  10.5170/CERN-2012-007} {\emph {\bibinfo {title} {{A Multi TeV Linear Collider
  based on CLIC technology: CLIC Conceptual Design Report}}}},\ \bibinfo {type}
  {Tech. Rep.}\ (\bibinfo  {institution} {CERN},\ \bibinfo {year}
  {2012})\BibitemShut {NoStop}%
\bibitem [{\citenamefont {Zennaro}\ \emph {et~al.}(2008)\citenamefont
  {Zennaro}, \citenamefont {Grudiev}, \citenamefont {Riddone}, \citenamefont
  {Samoshkin}, \citenamefont {Wuensch}, \citenamefont {Tantawi}, \citenamefont
  {Wang},\ and\ \citenamefont {Higo}}]{Zennaro2008DesignStructures}%
  \BibitemOpen
  \bibfield  {author} {\bibinfo {author} {\bibfnamefont {Riccardo}\
  \bibnamefont {Zennaro}}, \bibinfo {author} {\bibfnamefont {Alexej}\
  \bibnamefont {Grudiev}}, \bibinfo {author} {\bibfnamefont {G}~\bibnamefont
  {Riddone}}, \bibinfo {author} {\bibfnamefont {A}~\bibnamefont {Samoshkin}},
  \bibinfo {author} {\bibfnamefont {Walter}\ \bibnamefont {Wuensch}}, \bibinfo
  {author} {\bibfnamefont {Sami~G.}\ \bibnamefont {Tantawi}}, \bibinfo {author}
  {\bibfnamefont {J~W}\ \bibnamefont {Wang}}, \ and\ \bibinfo {author}
  {\bibfnamefont {Toshiyasu}\ \bibnamefont {Higo}},\ }\bibfield  {title}
  {\enquote {\bibinfo {title} {{Design and Fabrication of CLIC Test
  Structures}},}\ }in\ \href@noop {} {\emph {\bibinfo {booktitle} {Proceeding
  of LINAC 2008}}}\ (\bibinfo {address} {Victoria, BC, Canada},\ \bibinfo
  {year} {2008})\ pp.\ \bibinfo {pages} {533--535}\BibitemShut {NoStop}%
\bibitem [{\citenamefont {Descoeudres}\ \emph {et~al.}(2009)\citenamefont
  {Descoeudres}, \citenamefont {Ramsvik}, \citenamefont {Calatroni},
  \citenamefont {Taborelli},\ and\ \citenamefont
  {Wuensch}}]{Descoeudres2009DCVacuum}%
  \BibitemOpen
  \bibfield  {author} {\bibinfo {author} {\bibfnamefont {A}~\bibnamefont
  {Descoeudres}}, \bibinfo {author} {\bibfnamefont {T}~\bibnamefont {Ramsvik}},
  \bibinfo {author} {\bibfnamefont {Sergio}\ \bibnamefont {Calatroni}},
  \bibinfo {author} {\bibfnamefont {Mauro}\ \bibnamefont {Taborelli}}, \ and\
  \bibinfo {author} {\bibfnamefont {Walter}\ \bibnamefont {Wuensch}},\
  }\bibfield  {title} {\enquote {\bibinfo {title} {{DC breakdown conditioning
  and breakdown rate of metals and metallic alloys under ultrahigh vacuum}},}\
  }\href@noop {} {\bibfield  {journal} {\bibinfo  {journal} {Physical Review
  Special Topics-Accelerators and Beams}\ }\textbf {\bibinfo {volume} {12}},\
  \bibinfo {pages} {32001} (\bibinfo {year} {2009})}\BibitemShut {NoStop}%
\bibitem [{\citenamefont {Catalan-Lasheras}\ \emph {et~al.}(2014)\citenamefont
  {Catalan-Lasheras}, \citenamefont {Degiovanni}, \citenamefont {Doebert},
  \citenamefont {Farabolini}, \citenamefont {Kovermann}, \citenamefont
  {McMonagle}, \citenamefont {Rey}, \citenamefont {Syratchev}, \citenamefont
  {Timeo}, \citenamefont {Wuensch}, \citenamefont {Woolley},\ and\
  \citenamefont {Tagg}}]{Catalan-Lasheras2014ExperienceCERN}%
  \BibitemOpen
  \bibfield  {author} {\bibinfo {author} {\bibfnamefont {Nuria}\ \bibnamefont
  {Catalan-Lasheras}}, \bibinfo {author} {\bibfnamefont {Alberto}\ \bibnamefont
  {Degiovanni}}, \bibinfo {author} {\bibfnamefont {S}~\bibnamefont {Doebert}},
  \bibinfo {author} {\bibfnamefont {Wilfrid}\ \bibnamefont {Farabolini}},
  \bibinfo {author} {\bibfnamefont {Jan}\ \bibnamefont {Kovermann}}, \bibinfo
  {author} {\bibfnamefont {Gerard}\ \bibnamefont {McMonagle}}, \bibinfo
  {author} {\bibfnamefont {S}~\bibnamefont {Rey}}, \bibinfo {author}
  {\bibfnamefont {Igor}\ \bibnamefont {Syratchev}}, \bibinfo {author}
  {\bibfnamefont {L}~\bibnamefont {Timeo}}, \bibinfo {author} {\bibfnamefont
  {Walter}\ \bibnamefont {Wuensch}}, \bibinfo {author} {\bibfnamefont
  {Benjamin}\ \bibnamefont {Woolley}}, \ and\ \bibinfo {author} {\bibfnamefont
  {J}~\bibnamefont {Tagg}},\ }\bibfield  {title} {\enquote {\bibinfo {title}
  {{Experience Operating an X-Band High-Power Test Stand At CERN}},}\ }in\
  \href@noop {} {\emph {\bibinfo {booktitle} {Proceeding of IPAC2014}}}\
  (\bibinfo {year} {2014})\ pp.\ \bibinfo {pages} {2288--2290}\BibitemShut
  {NoStop}%
\bibitem [{\citenamefont {Catalan~Lasheras}\ \emph {et~al.}(2016)\citenamefont
  {Catalan~Lasheras}, \citenamefont {Solodko}, \citenamefont {Argyropoulos},
  \citenamefont {Woolley}, \citenamefont {Wuensch}, \citenamefont
  {Esperante~Pereira}, \citenamefont {Tagg}, \citenamefont {McMonagle},
  \citenamefont {Eymin},\ and\ \citenamefont
  {Syratchev}}]{CatalanLasheras2016CommissioningStand}%
  \BibitemOpen
  \bibfield  {author} {\bibinfo {author} {\bibfnamefont {Nuria}\ \bibnamefont
  {Catalan~Lasheras}}, \bibinfo {author} {\bibfnamefont {Anastasiya}\
  \bibnamefont {Solodko}}, \bibinfo {author} {\bibfnamefont {Theodoros}\
  \bibnamefont {Argyropoulos}}, \bibinfo {author} {\bibfnamefont {Benjamin}\
  \bibnamefont {Woolley}}, \bibinfo {author} {\bibfnamefont {Walter}\
  \bibnamefont {Wuensch}}, \bibinfo {author} {\bibfnamefont {Daniel}\
  \bibnamefont {Esperante~Pereira}}, \bibinfo {author} {\bibfnamefont
  {J}~\bibnamefont {Tagg}}, \bibinfo {author} {\bibfnamefont {Gerard}\
  \bibnamefont {McMonagle}}, \bibinfo {author} {\bibfnamefont {Cedric}\
  \bibnamefont {Eymin}}, \ and\ \bibinfo {author} {\bibfnamefont {Igor}\
  \bibnamefont {Syratchev}},\ }\bibfield  {title} {\enquote {\bibinfo {title}
  {{Commissioning of XBox-3: A very high capacity X-band test stand}},}\
  }\href@noop {} {\bibfield  {journal} {\bibinfo  {journal} {Proceedings of
  LINAC2016}\ } (\bibinfo {year} {2016})}\BibitemShut {NoStop}%
\bibitem [{\citenamefont {Volpi}\ \emph {et~al.}(2018)\citenamefont {Volpi},
  \citenamefont {Catalan-Lasheras}, \citenamefont {Grudiev}, \citenamefont
  {Lucas}, \citenamefont {McMonagle}, \citenamefont {Paszkiewicz},
  \citenamefont {Del}, \citenamefont {Romano}, \citenamefont {Syratchev},
  \citenamefont {Vnuchenko}, \citenamefont {Woolley}, \citenamefont {Wuensch},
  \citenamefont {Giansiracusa}, \citenamefont {Rassool}, \citenamefont
  {Serpico},\ and\ \citenamefont {Boland}}]{Volpi2018HighKlystrons}%
  \BibitemOpen
  \bibfield  {author} {\bibinfo {author} {\bibfnamefont {Matteo}\ \bibnamefont
  {Volpi}}, \bibinfo {author} {\bibfnamefont {Nuria}\ \bibnamefont
  {Catalan-Lasheras}}, \bibinfo {author} {\bibfnamefont {Alexej}\ \bibnamefont
  {Grudiev}}, \bibinfo {author} {\bibfnamefont {Thomas~Geoffrey}\ \bibnamefont
  {Lucas}}, \bibinfo {author} {\bibfnamefont {Gerard}\ \bibnamefont
  {McMonagle}}, \bibinfo {author} {\bibfnamefont {Jan}\ \bibnamefont
  {Paszkiewicz}}, \bibinfo {author} {\bibfnamefont {V}~\bibnamefont {Del}},
  \bibinfo {author} {\bibfnamefont {Pozo}\ \bibnamefont {Romano}}, \bibinfo
  {author} {\bibfnamefont {Igor}\ \bibnamefont {Syratchev}}, \bibinfo {author}
  {\bibfnamefont {Anna}\ \bibnamefont {Vnuchenko}}, \bibinfo {author}
  {\bibfnamefont {Benjamin}\ \bibnamefont {Woolley}}, \bibinfo {author}
  {\bibfnamefont {Walter}\ \bibnamefont {Wuensch}}, \bibinfo {author}
  {\bibfnamefont {P~J}\ \bibnamefont {Giansiracusa}}, \bibinfo {author}
  {\bibfnamefont {Roger~Paul}\ \bibnamefont {Rassool}}, \bibinfo {author}
  {\bibfnamefont {Claudio}\ \bibnamefont {Serpico}}, \ and\ \bibinfo {author}
  {\bibfnamefont {Mark~James}\ \bibnamefont {Boland}},\ }\bibfield  {title}
  {\enquote {\bibinfo {title} {{High Power and High Repetition Rate X-band
  Power Source Using Multiple Klystrons; High Power and High Repetition Rate
  X-band Power Source Using Multiple Klystrons}},}\ }\href {\doibase
  10.18429/JACoW-IPAC2018-THPMK104} {\  (\bibinfo {year} {2018}),\
  10.18429/JACoW-IPAC2018-THPMK104}\BibitemShut {NoStop}%
\bibitem [{\citenamefont {Shipman}(2014)}]{Shipman2014ExperimentalCLIC}%
  \BibitemOpen
  \bibfield  {author} {\bibinfo {author} {\bibfnamefont {Nicholas}\
  \bibnamefont {Shipman}},\ }\emph {\bibinfo {title} {{Experimental study of DC
  vacuum breakdown and application to high-gradient accelerating structures for
  CLIC}}},\ \href@noop {} {Ph.D. thesis} (\bibinfo {year} {2014})\BibitemShut
  {NoStop}%
\bibitem [{\citenamefont {Woolley}(2015)}]{Woolley2015HighCavity}%
  \BibitemOpen
  \bibfield  {author} {\bibinfo {author} {\bibfnamefont {Benjamin}\
  \bibnamefont {Woolley}},\ }\emph {\bibinfo {title} {{High power X-band RF
  test stand development and high power testing of the CLIC crab cavity}}},\
  \href
  {https://ezp.lib.unimelb.edu.au/login?url=https://search.ebscohost.com/login.aspx?direct=true&db=edsble&AN=edsble.677264&site=eds-live&scope=site}
  {Ph.D. thesis} (\bibinfo {year} {2015})\BibitemShut {NoStop}%
\bibitem [{201(2015)}]{2015ManualEPULSUS-FPM1-10}%
  \BibitemOpen
  \href@noop {} {\enquote {\bibinfo {title} {{Manual EPULSUS-FPM1-10}},}\ }
  (\bibinfo {year} {2015})\BibitemShut {NoStop}%
\bibitem [{\citenamefont {Redondo}\ \emph {et~al.}(2017)\citenamefont
  {Redondo}, \citenamefont {Kandratsyeu}, \citenamefont {Barnes}, \citenamefont
  {Calatroni},\ and\ \citenamefont {Wuensch}}]{Redondo2017Solid-stateStudies}%
  \BibitemOpen
  \bibfield  {author} {\bibinfo {author} {\bibfnamefont {L.~M.}\ \bibnamefont
  {Redondo}}, \bibinfo {author} {\bibfnamefont {Aleh}\ \bibnamefont
  {Kandratsyeu}}, \bibinfo {author} {\bibfnamefont {M.~J.}\ \bibnamefont
  {Barnes}}, \bibinfo {author} {\bibfnamefont {Sergio}\ \bibnamefont
  {Calatroni}}, \ and\ \bibinfo {author} {\bibfnamefont {Walter}\ \bibnamefont
  {Wuensch}},\ }\bibfield  {title} {\enquote {\bibinfo {title} {{Solid-state
  Marx generator for the compact linear collider breakdown studies}},}\ }in\
  \href {\doibase 10.1109/IPMHVC.2016.8012824} {\emph {\bibinfo {booktitle}
  {2016 IEEE International Power Modulator and High Voltage Conference, IPMHVC
  2016}}}\ (\bibinfo {year} {2017})\BibitemShut {NoStop}%
\bibitem [{\citenamefont {J{\^{u}}ttner}(2001)}]{Juttner2001CathodeArcs}%
  \BibitemOpen
  \bibfield  {author} {\bibinfo {author} {\bibfnamefont {Burkhard}\
  \bibnamefont {J{\^{u}}ttner}},\ }\bibfield  {title} {\enquote {\bibinfo
  {title} {{Cathode spots of electric arcs}},}\ }\href@noop {} {\bibfield
  {journal} {\bibinfo  {journal} {J. Phys. D: Appl. Phys}\ }\textbf {\bibinfo
  {volume} {34}},\ \bibinfo {pages} {103--123} (\bibinfo {year}
  {2001})}\BibitemShut {NoStop}%
\bibitem [{\citenamefont {Shipman}\ \emph {et~al.}(2012)\citenamefont
  {Shipman}, \citenamefont {Calatroni}, \citenamefont {Wuensch},\ and\
  \citenamefont {Jones}}]{Shipman2012MeasurementTime}%
  \BibitemOpen
  \bibfield  {author} {\bibinfo {author} {\bibfnamefont {Nicholas}\
  \bibnamefont {Shipman}}, \bibinfo {author} {\bibfnamefont {Sergio}\
  \bibnamefont {Calatroni}}, \bibinfo {author} {\bibfnamefont {Walter}\
  \bibnamefont {Wuensch}}, \ and\ \bibinfo {author} {\bibfnamefont {R.~M.}\
  \bibnamefont {Jones}},\ }\bibfield  {title} {\enquote {\bibinfo {title}
  {{Measurement of the Dynamic Response of the CERN DC Spark System and
  Preliminary Estimates of the Breakdown Turn-on Time}},}\ }in\ \href@noop {}
  {\emph {\bibinfo {booktitle} {IPAC2012}}}\ (\bibinfo  {publisher} {IEEE},\
  \bibinfo {address} {New Orleans, Louisiana, USA},\ \bibinfo {year}
  {2012})\BibitemShut {NoStop}%
\bibitem [{\citenamefont {Cox}(1974)}]{Cox1974VariationVacuo}%
  \BibitemOpen
  \bibfield  {author} {\bibinfo {author} {\bibfnamefont {B~M}\ \bibnamefont
  {Cox}},\ }\bibfield  {title} {\enquote {\bibinfo {title} {{Variation of the
  critical breakdown field between copper electrodes in vacuo}},}\ }\href@noop
  {} {\bibfield  {journal} {\bibinfo  {journal} {Journal of Physics D: Applied
  Physics}\ }\textbf {\bibinfo {volume} {7}},\ \bibinfo {pages} {143} (\bibinfo
  {year} {1974})}\BibitemShut {NoStop}%
\bibitem [{\citenamefont {Korsb{\"{a}}ck}\ \emph {et~al.}(2019)\citenamefont
  {Korsb{\"{a}}ck}, \citenamefont {Morales}, \citenamefont {Profatilova},
  \citenamefont {Rodriquez~Castro}, \citenamefont {Wuensch}, \citenamefont
  {Calatroni},\ and\ \citenamefont {Ahlgren}}]{Korsback2019VacuumSystem}%
  \BibitemOpen
  \bibfield  {author} {\bibinfo {author} {\bibfnamefont {Anders}\ \bibnamefont
  {Korsb{\"{a}}ck}}, \bibinfo {author} {\bibfnamefont {Laura~Mercadé}\
  \bibnamefont {Morales}}, \bibinfo {author} {\bibfnamefont {Iaroslava}\
  \bibnamefont {Profatilova}}, \bibinfo {author} {\bibfnamefont {Enrique}\
  \bibnamefont {Rodriquez~Castro}}, \bibinfo {author} {\bibfnamefont {Walter}\
  \bibnamefont {Wuensch}}, \bibinfo {author} {\bibfnamefont {Sergio}\
  \bibnamefont {Calatroni}}, \ and\ \bibinfo {author} {\bibfnamefont {Tommy}\
  \bibnamefont {Ahlgren}},\ }\bibfield  {title} {\enquote {\bibinfo {title}
  {{Vacuum electrical breakdown conditioning study in a parallel plate
  electrode pulsed DC system}},}\ }\href@noop {} {\bibfield  {journal}
  {\bibinfo  {journal} {Arxiv}\ }\textbf {\bibinfo {volume} {1905.03996}}
  (\bibinfo {year} {2019})}\BibitemShut {NoStop}%
\bibitem [{\citenamefont {Gudkov}(2014)}]{Gudkov2014CDD:Disk}%
  \BibitemOpen
  \bibfield  {author} {\bibinfo {author} {\bibfnamefont {Dmitry}\ \bibnamefont
  {Gudkov}},\ }\href {https://edms5.cern.ch/document/1162463/AA} {\enquote
  {\bibinfo {title} {{CDD: Drawings, Fixed Gap System Anode / Cathode Disk}},}\
  } (\bibinfo {year} {2014})\BibitemShut {NoStop}%
\bibitem [{\citenamefont {{ASTM
  International}}(2013)}]{ASTMInternational2013ASTMSize}%
  \BibitemOpen
  \bibfield  {author} {\bibinfo {author} {\bibnamefont {{ASTM
  International}}},\ }\href {\doibase 10.1520/E0112-10} {\enquote {\bibinfo
  {title} {{ASTM E112-13, Standard Test Methods for Determining Average Grain
  Size}},}\ } (\bibinfo {year} {2013})\BibitemShut {NoStop}%
\bibitem [{\citenamefont {Profatilova}\ \emph {et~al.}(2019)\citenamefont
  {Profatilova}, \citenamefont {Stragier}, \citenamefont {Calatroni},
  \citenamefont {Kandratsyeu}, \citenamefont {Rodriquez~Castro},\ and\
  \citenamefont {Wuensch}}]{Profatilova2019BreakdownSystem}%
  \BibitemOpen
  \bibfield  {author} {\bibinfo {author} {\bibfnamefont {Iaroslava}\
  \bibnamefont {Profatilova}}, \bibinfo {author} {\bibfnamefont {Xavier}\
  \bibnamefont {Stragier}}, \bibinfo {author} {\bibfnamefont {Sergio}\
  \bibnamefont {Calatroni}}, \bibinfo {author} {\bibfnamefont {Aleh}\
  \bibnamefont {Kandratsyeu}}, \bibinfo {author} {\bibfnamefont {Enrique}\
  \bibnamefont {Rodriquez~Castro}}, \ and\ \bibinfo {author} {\bibfnamefont
  {Walter}\ \bibnamefont {Wuensch}},\ }\bibfield  {title} {\enquote {\bibinfo
  {title} {{Breakdown localisation in a pulsed DC electrode system}},}\ }\href
  {\doibase 10.1016/j.nima.2019.163079} {\bibfield  {journal} {\bibinfo
  {journal} {Nuclear Inst. and Methods in Physics Research, A}\ ,\ \bibinfo
  {pages} {163079}} (\bibinfo {year} {2019})}\BibitemShut {NoStop}%
\bibitem [{\citenamefont {Parker}(2002)}]{Parker2002ElectricCapacitor}%
  \BibitemOpen
  \bibfield  {author} {\bibinfo {author} {\bibfnamefont {G.~W.}\ \bibnamefont
  {Parker}},\ }\bibfield  {title} {\enquote {\bibinfo {title} {{Electric field
  outside a parallel plate capacitor}},}\ }\href {\doibase 10.1119/1.1463738}
  {\bibfield  {journal} {\bibinfo  {journal} {American Journal of Physics}\
  }\textbf {\bibinfo {volume} {70}},\ \bibinfo {pages} {502--507} (\bibinfo
  {year} {2002})}\BibitemShut {NoStop}%
\bibitem [{\citenamefont {Catal{\'{a}}n~Izquierdo}\ \emph
  {et~al.}(2017)\citenamefont {Catal{\'{a}}n~Izquierdo}, \citenamefont
  {Bueno~Barrachina}, \citenamefont {Ca{\~{n}}as~Pe{\~{n}}uelas},\ and\
  \citenamefont
  {Cavall{\'{e}}~Ses{\'{e}}}}]{CatalanIzquierdo2017CapacitanceAnalysis}%
  \BibitemOpen
  \bibfield  {author} {\bibinfo {author} {\bibfnamefont {S.}~\bibnamefont
  {Catal{\'{a}}n~Izquierdo}}, \bibinfo {author} {\bibfnamefont {José~M.}\
  \bibnamefont {Bueno~Barrachina}}, \bibinfo {author} {\bibfnamefont
  {César~S.}\ \bibnamefont {Ca{\~{n}}as~Pe{\~{n}}uelas}}, \ and\ \bibinfo
  {author} {\bibfnamefont {Francisco}\ \bibnamefont
  {Cavall{\'{e}}~Ses{\'{e}}}},\ }\bibfield  {title} {\enquote {\bibinfo {title}
  {{Capacitance evaluation on parallel-plate capacitors by means of finite
  element analysis}},}\ }\href {\doibase 10.24084/repqj07.451} {\bibfield
  {journal} {\bibinfo  {journal} {Renewable Energy and Power Quality Journal}\
  }\textbf {\bibinfo {volume} {1}},\ \bibinfo {pages} {613--616} (\bibinfo
  {year} {2017})}\BibitemShut {NoStop}%
\bibitem [{\citenamefont {Kassamakov}\ \emph {et~al.}(2017)\citenamefont
  {Kassamakov}, \citenamefont {Lecler}, \citenamefont {Nolvi}, \citenamefont
  {Leong-Ho{\"{i}}}, \citenamefont {Montgomery},\ and\ \citenamefont
  {H{\ae}ggstr{\"{o}}m}}]{Kassamakov20173DInterferometry}%
  \BibitemOpen
  \bibfield  {author} {\bibinfo {author} {\bibfnamefont {Ivan}\ \bibnamefont
  {Kassamakov}}, \bibinfo {author} {\bibfnamefont {Sylvain}\ \bibnamefont
  {Lecler}}, \bibinfo {author} {\bibfnamefont {Anton}\ \bibnamefont {Nolvi}},
  \bibinfo {author} {\bibfnamefont {Audrey}\ \bibnamefont {Leong-Ho{\"{i}}}},
  \bibinfo {author} {\bibfnamefont {Paul}\ \bibnamefont {Montgomery}}, \ and\
  \bibinfo {author} {\bibfnamefont {Edward}\ \bibnamefont
  {H{\ae}ggstr{\"{o}}m}},\ }\bibfield  {title} {\enquote {\bibinfo {title} {{3D
  super-resolution optical profiling using microsphere enhanced Mirau
  interferometry}},}\ }\href@noop {} {\bibfield  {journal} {\bibinfo  {journal}
  {Scientific reports}\ }\textbf {\bibinfo {volume} {7}},\ \bibinfo {pages}
  {3683} (\bibinfo {year} {2017})}\BibitemShut {NoStop}%
\bibitem [{\citenamefont {Wang}\ and\ \citenamefont
  {Loew}(1989)}]{Wang1989RfStructures}%
  \BibitemOpen
  \bibfield  {author} {\bibinfo {author} {\bibfnamefont {J~W}\ \bibnamefont
  {Wang}}\ and\ \bibinfo {author} {\bibfnamefont {G~A}\ \bibnamefont {Loew}},\
  }\bibfield  {title} {\enquote {\bibinfo {title} {{Rf breakdown studies in
  copper electron linac structures}},}\ }in\ \href@noop {} {\emph {\bibinfo
  {booktitle} {Proceedings of the 1989 IEEE Particle Accelerator
  Conference,.'Accelerator Science and Technology}}}\ (\bibinfo {year} {1989})\
  pp.\ \bibinfo {pages} {1137--1139}\BibitemShut {NoStop}%
\bibitem [{\citenamefont {Zennaro}\ \emph {et~al.}(2017)\citenamefont
  {Zennaro}, \citenamefont {Garvey}, \citenamefont {Argyropoulos},
  \citenamefont {Wegner}, \citenamefont {Grudiev}, \citenamefont {Wuensch},
  \citenamefont {Esperante~Pereira}, \citenamefont {McMonagle}, \citenamefont
  {Woolley},\ and\ \citenamefont {Lucas}}]{Zennaro2017HighCLIC}%
  \BibitemOpen
  \bibfield  {author} {\bibinfo {author} {\bibfnamefont {Riccardo}\
  \bibnamefont {Zennaro}}, \bibinfo {author} {\bibfnamefont {Terence}\
  \bibnamefont {Garvey}}, \bibinfo {author} {\bibfnamefont {Theodoros}\
  \bibnamefont {Argyropoulos}}, \bibinfo {author} {\bibfnamefont {Rolf}\
  \bibnamefont {Wegner}}, \bibinfo {author} {\bibfnamefont {Alexej}\
  \bibnamefont {Grudiev}}, \bibinfo {author} {\bibfnamefont {Walter}\
  \bibnamefont {Wuensch}}, \bibinfo {author} {\bibfnamefont {Daniel}\
  \bibnamefont {Esperante~Pereira}}, \bibinfo {author} {\bibfnamefont {Gerard}\
  \bibnamefont {McMonagle}}, \bibinfo {author} {\bibfnamefont {Benjamin}\
  \bibnamefont {Woolley}}, \ and\ \bibinfo {author} {\bibfnamefont {Thomas}\
  \bibnamefont {Lucas}},\ }\bibfield  {title} {\enquote {\bibinfo {title}
  {{High power tests of a prototype X-band accelerating structure for CLIC}},}\
  }\href@noop {} {\bibfield  {journal} {\bibinfo  {journal} {CLIC Note}\
  }\textbf {\bibinfo {volume} {1134}} (\bibinfo {year} {2017})}\BibitemShut
  {NoStop}%
\bibitem [{\citenamefont {Wuensch}\ \emph {et~al.}(2017)\citenamefont
  {Wuensch}, \citenamefont {Degiovanni}, \citenamefont {Calatroni},
  \citenamefont {Korsb{\"{a}}ck}, \citenamefont {Djurabekova}, \citenamefont
  {Rajam{\"{a}}ki},\ and\ \citenamefont
  {Navarro}}]{Wuensch2017StatisticsRegime}%
  \BibitemOpen
  \bibfield  {author} {\bibinfo {author} {\bibfnamefont {Walter}\ \bibnamefont
  {Wuensch}}, \bibinfo {author} {\bibfnamefont {Alberto}\ \bibnamefont
  {Degiovanni}}, \bibinfo {author} {\bibfnamefont {Sergio}\ \bibnamefont
  {Calatroni}}, \bibinfo {author} {\bibfnamefont {Anders}\ \bibnamefont
  {Korsb{\"{a}}ck}}, \bibinfo {author} {\bibfnamefont {Flyura}\ \bibnamefont
  {Djurabekova}}, \bibinfo {author} {\bibfnamefont {Robin}\ \bibnamefont
  {Rajam{\"{a}}ki}}, \ and\ \bibinfo {author} {\bibfnamefont {Jorge~Giner}\
  \bibnamefont {Navarro}},\ }\bibfield  {title} {\enquote {\bibinfo {title}
  {{Statistics of vacuum breakdown in the high-gradient and low-rate
  regime}},}\ }\href {\doibase 10.1103/PhysRevAccelBeams.20.011007} {\bibfield
  {journal} {\bibinfo  {journal} {Physical Review Accelerators and Beams}\
  }\textbf {\bibinfo {volume} {20}},\ \bibinfo {pages} {11007} (\bibinfo {year}
  {2017})}\BibitemShut {NoStop}%
\bibitem [{\citenamefont {Papanikolaou}\ \emph {et~al.}(2011)\citenamefont
  {Papanikolaou}, \citenamefont {Bohn}, \citenamefont {Sommer}, \citenamefont
  {Durin}, \citenamefont {Zapperi},\ and\ \citenamefont
  {Sethna}}]{Papanikolaou2011UniversalityShape}%
  \BibitemOpen
  \bibfield  {author} {\bibinfo {author} {\bibfnamefont {Stefanos}\
  \bibnamefont {Papanikolaou}}, \bibinfo {author} {\bibfnamefont {Felipe}\
  \bibnamefont {Bohn}}, \bibinfo {author} {\bibfnamefont {Rubem~Luis}\
  \bibnamefont {Sommer}}, \bibinfo {author} {\bibfnamefont {Gianfranco}\
  \bibnamefont {Durin}}, \bibinfo {author} {\bibfnamefont {Stefano}\
  \bibnamefont {Zapperi}}, \ and\ \bibinfo {author} {\bibfnamefont {James~P}\
  \bibnamefont {Sethna}},\ }\bibfield  {title} {\enquote {\bibinfo {title}
  {{Universality beyond power laws and the average avalanche shape}},}\
  }\href@noop {} {\bibfield  {journal} {\bibinfo  {journal} {Nature Physics}\
  }\textbf {\bibinfo {volume} {7}},\ \bibinfo {pages} {316} (\bibinfo {year}
  {2011})}\BibitemShut {NoStop}%
\bibitem [{\citenamefont {Chrzan}\ and\ \citenamefont
  {Mills}(1994)}]{Chrzan1994CriticalityCompounds}%
  \BibitemOpen
  \bibfield  {author} {\bibinfo {author} {\bibfnamefont {D~C}\ \bibnamefont
  {Chrzan}}\ and\ \bibinfo {author} {\bibfnamefont {M~J}\ \bibnamefont
  {Mills}},\ }\bibfield  {title} {\enquote {\bibinfo {title} {{Criticality in
  the plastic deformation of L 1 2 intermetallic compounds}},}\ }\href@noop {}
  {\bibfield  {journal} {\bibinfo  {journal} {Physical Review B}\ }\textbf
  {\bibinfo {volume} {50}},\ \bibinfo {pages} {30} (\bibinfo {year}
  {1994})}\BibitemShut {NoStop}%
\bibitem [{\citenamefont {Davies}(1973)}]{Davies1973TheReview}%
  \BibitemOpen
  \bibfield  {author} {\bibinfo {author} {\bibfnamefont {D.~Kenneth}\
  \bibnamefont {Davies}},\ }\bibfield  {title} {\enquote {\bibinfo {title}
  {{The initiation of electrical breakdown in vacuum—a review}},}\
  }\href@noop {} {\bibfield  {journal} {\bibinfo  {journal} {Journal of Vacuum
  Science and Technology}\ }\textbf {\bibinfo {volume} {10}},\ \bibinfo {pages}
  {115--121} (\bibinfo {year} {1973})}\BibitemShut {NoStop}%
\bibitem [{\citenamefont {Davies}\ and\ \citenamefont
  {Biondi}(1966)}]{Davies1966VacuumElectrodes}%
  \BibitemOpen
  \bibfield  {author} {\bibinfo {author} {\bibfnamefont {D.~Kenneth}\
  \bibnamefont {Davies}}\ and\ \bibinfo {author} {\bibfnamefont {Manfred~A.}\
  \bibnamefont {Biondi}},\ }\bibfield  {title} {\enquote {\bibinfo {title}
  {{Vacuum electrical breakdown between plane-parallel copper electrodes}},}\
  }\href {\doibase 10.1063/1.1703148} {\bibfield  {journal} {\bibinfo
  {journal} {Journal of Applied Physics}\ }\textbf {\bibinfo {volume} {37}},\
  \bibinfo {pages} {2969--2977} (\bibinfo {year} {1966})}\BibitemShut {NoStop}%
\bibitem [{\citenamefont {Adolphsen}\ \emph {et~al.}(2001)\citenamefont
  {Adolphsen}, \citenamefont {Baumgartner}, \citenamefont {Jobe}, \citenamefont
  {Le~Pimpec}, \citenamefont {Loewen}, \citenamefont {McCormick}, \citenamefont
  {Ross}, \citenamefont {Smith}, \citenamefont {Wang},\ and\ \citenamefont
  {Higo}}]{Adolphsen2001ProcessingNLCTA}%
  \BibitemOpen
  \bibfield  {author} {\bibinfo {author} {\bibfnamefont {C.}~\bibnamefont
  {Adolphsen}}, \bibinfo {author} {\bibfnamefont {W}~\bibnamefont
  {Baumgartner}}, \bibinfo {author} {\bibfnamefont {K}~\bibnamefont {Jobe}},
  \bibinfo {author} {\bibfnamefont {F}~\bibnamefont {Le~Pimpec}}, \bibinfo
  {author} {\bibfnamefont {R}~\bibnamefont {Loewen}}, \bibinfo {author}
  {\bibfnamefont {Doug}\ \bibnamefont {McCormick}}, \bibinfo {author}
  {\bibfnamefont {M}~\bibnamefont {Ross}}, \bibinfo {author} {\bibfnamefont
  {T}~\bibnamefont {Smith}}, \bibinfo {author} {\bibfnamefont {J~W}\
  \bibnamefont {Wang}}, \ and\ \bibinfo {author} {\bibfnamefont {Toshiyasu}\
  \bibnamefont {Higo}},\ }\bibfield  {title} {\enquote {\bibinfo {title}
  {{Processing studies of X-band accelerator structures at the NLCTA}},}\ }in\
  \href {\doibase 10.1109/pac.2001.987546} {\emph {\bibinfo {booktitle}
  {Proceeding of the 2001 Particle Accelerator Conference}}},\ \bibinfo {series
  and number} {\bibinfo {number} {1}}\ (\bibinfo {address} {Chicago,
  Illinois},\ \bibinfo {year} {2001})\ pp.\ \bibinfo {pages}
  {478--480}\BibitemShut {NoStop}%
\bibitem [{\citenamefont {Dal~Forno}\ \emph {et~al.}(2016)\citenamefont
  {Dal~Forno}, \citenamefont {Dolgashev}, \citenamefont {Bowden}, \citenamefont
  {Clarke}, \citenamefont {Hogan}, \citenamefont {McCormick}, \citenamefont
  {Novokhatski}, \citenamefont {Spataro}, \citenamefont {Weathersby},\ and\
  \citenamefont {Tantawi}}]{DalForno2016RfStructures}%
  \BibitemOpen
  \bibfield  {author} {\bibinfo {author} {\bibfnamefont {Massimo}\ \bibnamefont
  {Dal~Forno}}, \bibinfo {author} {\bibfnamefont {Valery~A.}\ \bibnamefont
  {Dolgashev}}, \bibinfo {author} {\bibfnamefont {Gordon}\ \bibnamefont
  {Bowden}}, \bibinfo {author} {\bibfnamefont {Christine}\ \bibnamefont
  {Clarke}}, \bibinfo {author} {\bibfnamefont {Mark}\ \bibnamefont {Hogan}},
  \bibinfo {author} {\bibfnamefont {Doug}\ \bibnamefont {McCormick}}, \bibinfo
  {author} {\bibfnamefont {Alexander}\ \bibnamefont {Novokhatski}}, \bibinfo
  {author} {\bibfnamefont {Bruno}\ \bibnamefont {Spataro}}, \bibinfo {author}
  {\bibfnamefont {Stephen}\ \bibnamefont {Weathersby}}, \ and\ \bibinfo
  {author} {\bibfnamefont {Sami~G.}\ \bibnamefont {Tantawi}},\ }\bibfield
  {title} {\enquote {\bibinfo {title} {{rf breakdown tests of mm-wave metallic
  accelerating structures}},}\ }\href {\doibase
  10.1103/PhysRevAccelBeams.19.011301} {\bibfield  {journal} {\bibinfo
  {journal} {Physical Review Accelerators and Beams}\ }\textbf {\bibinfo
  {volume} {19}} (\bibinfo {year} {2016}),\
  10.1103/PhysRevAccelBeams.19.011301}\BibitemShut {NoStop}%
\bibitem [{\citenamefont {Degiovanni}\ \emph {et~al.}(2016)\citenamefont
  {Degiovanni}, \citenamefont {Wuensch},\ and\ \citenamefont
  {Navarro}}]{Degiovanni2016ComparisonStructures}%
  \BibitemOpen
  \bibfield  {author} {\bibinfo {author} {\bibfnamefont {Alberto}\ \bibnamefont
  {Degiovanni}}, \bibinfo {author} {\bibfnamefont {Walter}\ \bibnamefont
  {Wuensch}}, \ and\ \bibinfo {author} {\bibfnamefont {Jorge~Giner}\
  \bibnamefont {Navarro}},\ }\bibfield  {title} {\enquote {\bibinfo {title}
  {{Comparison of the conditioning of high gradient accelerating
  structures}},}\ }\href {\doibase 10.1103/PhysRevAccelBeams.19.032001}
  {\bibfield  {journal} {\bibinfo  {journal} {Physical Review Accelerators and
  Beams}\ } (\bibinfo {year} {2016}),\
  10.1103/PhysRevAccelBeams.19.032001}\BibitemShut {NoStop}%
\bibitem [{\citenamefont {Mesyats}(2013)}]{Mesyats2013EctonArc}%
  \BibitemOpen
  \bibfield  {author} {\bibinfo {author} {\bibfnamefont {Gennady~A.}\
  \bibnamefont {Mesyats}},\ }\bibfield  {title} {\enquote {\bibinfo {title}
  {{Ecton mechanism of the cathode spot phenomena in a vacuum Arc}},}\ }\href
  {\doibase 10.1109/TPS.2013.2247064} {\bibfield  {journal} {\bibinfo
  {journal} {IEEE Transactions on Plasma Science}\ }\textbf {\bibinfo {volume}
  {41}},\ \bibinfo {pages} {676--694} (\bibinfo {year} {2013})}\BibitemShut
  {NoStop}%
\bibitem [{\citenamefont {Hull}\ and\ \citenamefont
  {Bacon}(2011)}]{Hull2011IntroductionDislocations}%
  \BibitemOpen
  \bibfield  {author} {\bibinfo {author} {\bibfnamefont {Derek}\ \bibnamefont
  {Hull}}\ and\ \bibinfo {author} {\bibfnamefont {David~J}\ \bibnamefont
  {Bacon}},\ }\href@noop {} {\emph {\bibinfo {title} {{Introduction to
  dislocations}}}}\ (\bibinfo  {publisher} {Butterworth-Heinemann},\ \bibinfo
  {year} {2011})\BibitemShut {NoStop}%
\bibitem [{\citenamefont {Pohjonen}\ \emph {et~al.}(2012)\citenamefont
  {Pohjonen}, \citenamefont {Djurabekova}, \citenamefont {Kuronen},
  \citenamefont {Fitzgerald},\ and\ \citenamefont
  {Nordlund}}]{Pohjonen2012AnalyticalStress}%
  \BibitemOpen
  \bibfield  {author} {\bibinfo {author} {\bibfnamefont {Aarne~S.}\
  \bibnamefont {Pohjonen}}, \bibinfo {author} {\bibfnamefont {Flyura}\
  \bibnamefont {Djurabekova}}, \bibinfo {author} {\bibfnamefont {Antti}\
  \bibnamefont {Kuronen}}, \bibinfo {author} {\bibfnamefont {Steven~P.}\
  \bibnamefont {Fitzgerald}}, \ and\ \bibinfo {author} {\bibfnamefont {Kai}\
  \bibnamefont {Nordlund}},\ }\bibfield  {title} {\enquote {\bibinfo {title}
  {{Analytical model of dislocation nucleation on a near-surface void under
  tensile surface stress}},}\ }\href {\doibase 10.1080/14786435.2012.700415}
  {\bibfield  {journal} {\bibinfo  {journal} {Philosophical Magazine}\ }
  (\bibinfo {year} {2012}),\ 10.1080/14786435.2012.700415}\BibitemShut
  {NoStop}%
\bibitem [{\citenamefont {Cahill}\ \emph {et~al.}(2018)\citenamefont {Cahill},
  \citenamefont {Rosenzweig}, \citenamefont {Dolgashev}, \citenamefont
  {Tantawi},\ and\ \citenamefont {Weathersby}}]{Cahill2018HighCavities}%
  \BibitemOpen
  \bibfield  {author} {\bibinfo {author} {\bibfnamefont {A.~D.}\ \bibnamefont
  {Cahill}}, \bibinfo {author} {\bibfnamefont {J.~B.}\ \bibnamefont
  {Rosenzweig}}, \bibinfo {author} {\bibfnamefont {Valery~A.}\ \bibnamefont
  {Dolgashev}}, \bibinfo {author} {\bibfnamefont {Sami~G.}\ \bibnamefont
  {Tantawi}}, \ and\ \bibinfo {author} {\bibfnamefont {Stephen}\ \bibnamefont
  {Weathersby}},\ }\bibfield  {title} {\enquote {\bibinfo {title} {{High
  gradient experiments with X -band cryogenic copper accelerating cavities}},}\
  }\href {\doibase 10.1103/PhysRevAccelBeams.21.102002} {\bibfield  {journal}
  {\bibinfo  {journal} {Physical Review Accelerators and Beams}\ }\textbf
  {\bibinfo {volume} {21}},\ \bibinfo {pages} {102002} (\bibinfo {year}
  {2018})}\BibitemShut {NoStop}%
\bibitem [{\citenamefont {Abe}\ \emph {et~al.}(2018)\citenamefont {Abe},
  \citenamefont {Kageyama}, \citenamefont {Sakai}, \citenamefont {Takeuchi},\
  and\ \citenamefont {Yoshino}}]{Abe2018DirectCavity}%
  \BibitemOpen
  \bibfield  {author} {\bibinfo {author} {\bibfnamefont {Tetsuo}\ \bibnamefont
  {Abe}}, \bibinfo {author} {\bibfnamefont {Tatsuya}\ \bibnamefont {Kageyama}},
  \bibinfo {author} {\bibfnamefont {Hiroshi}\ \bibnamefont {Sakai}}, \bibinfo
  {author} {\bibfnamefont {Yasunao}\ \bibnamefont {Takeuchi}}, \ and\ \bibinfo
  {author} {\bibfnamefont {Kazuo}\ \bibnamefont {Yoshino}},\ }\bibfield
  {title} {\enquote {\bibinfo {title} {{Direct observation of breakdown trigger
  seeds in a normal-conducting rf accelerating cavity}},}\ }\href {\doibase
  10.1103/PhysRevAccelBeams.21.122002} {\bibfield  {journal} {\bibinfo
  {journal} {Physical Review Accelerators and Beams}\ }\textbf {\bibinfo
  {volume} {21}} (\bibinfo {year} {2018}),\
  10.1103/PhysRevAccelBeams.21.122002}\BibitemShut {NoStop}%
\end{thebibliography}%
\end{document}